\documentclass[a4wide,12pt]{article}
\usepackage[margin=1.2in]{geometry}
\usepackage[utf8]{inputenc}
\usepackage [english] {babel} 
\usepackage{amsmath,amssymb,latexsym}
\usepackage{slashed}
\usepackage{graphicx}
\usepackage{epstopdf}
\usepackage{relsize}
\usepackage{young}
\usepackage{mathtools}
\usepackage[sorting=none, backend=bibtex]{biblatex}
\bibliography{ref}
\DeclareGraphicsExtensions{.png}

\usepackage{hyperref}
\usepackage{pict2e}
\def \sec{\begin{section}}
\def \esec{\end{section}}

\DeclarePairedDelimiter\floor{\lfloor}{\rfloor}

\def \la {\lambda}

\def \Om {\Omega}

\def \ep {\epsilon}

\def \pr {\partial}

\def \ra {\rightarrow}

\def \beq { \begin{equation}}

\def \eeq {\end{equation}}

\newcommand\const{\operatorname{const}}

\def \l {\left(}

\def \r {\right)}

\def \ll {\langle}

\def \rr {\rangle}

\def \homf {H^{(n,m)}_\square}

\def \homfnk {H^{n,nk+1}_\square}

\def \sunt {\mathcal{N}=2}

\def \suno {\mathcal{N}=1}

\def \sunf {\mathcal{N}=4}

\def \st {S^1_\beta \times \mathbb{R}^4_\Omega}

\def \at {\biggl{\vert}}

\begin{document}

\begin{titlepage}

\begin{center}

{  \Large \bf The Condensate from Torus Knots }

\end{center}

\vspace{1mm}

\begin{center}

{\large

  A.~Gorsky$^{\,2,3}$,    A.~Milekhin$^{\,1,2,3,4}$\ and N.~Sopenko$^{\,1,2}$\ }

\vspace{3mm}

$^1$Institute of Theoretical and Experimental Physics, B.Cheryomushkinskaya 25, Moscow 117218, Russia \\
$^2$Moscow Institute of Physics and Technology, Dolgoprudny 141700, Russia \\
$^3$ Institute for Information Transmission Problems of Russian Academy of Science, B. Karetnyi 19, Moscow 127051, Russia\\
$^4$ Department of Physics, Princeton University, Princeton, NJ 08544 \\

\vspace{1cm} 

gorsky@itep.ru, milekhin@itep.ru, niksopenko@gmail.com

\end{center}

\vspace{1cm}


\begin{center}

{\large \bf Abstract}

\end{center}

We discuss  recently formulated instanton-torus knot duality in
$\Omega$-deformed 5D SQED on $\mathbb{R}^4 \times S^1$ focusing at the microscopic aspects  of 
the condensate  formation in the instanton ensemble. Using the chain of 
dualities and geometric transitions we embed the
SQED with a surface defect into the $SU(2)$ SQCD  with $N_f=4$ and identify the  numbers $(n,m)$ of
the torus $T_{n,m}$ knot as  instanton charge and  electric charge.
The HOMFLY torus  knot invariants in the fundamental representation provide   entropic factor in the condensate 
of the massless flavor counting the degeneracy of the instanton--W-boson web with instanton 
and electric numbers $(n,m)$ but different  spin and flavor content. Using the inverse geometrical transition we  explain how our approach is related  to the evaluation
of the HOMFLY invariants in terms of Wilson loop in 3d CS theory.
The reduction to   4D theory is briefly considered and some analogy with 
baryon vertex is conjectured.

\end{titlepage}

\tableofcontents

\newpage

\section{Introduction}

The clarification of the microscopic mechanism behind the formation of the
condensates is the challenging problem in a quantum field theory. Usually 
this question is  substituted  by a kind of  a  mean field analysis.
However  in some cases it is possible to recognize
that particular non-perturbative configuration or ensembles of the non-perturbative
configurations are responsible for the condensate formation. The familiar example
is  evaluation of the gluino condensate in the SYM theory in terms
of the gluino zero modes in the instanton background \cite{nsvz}. The issue
is quite subtle since for instance in SU(2) SYM theory the instanton configuration
saturates only the topological correlator and additional clusterization argument
has to be applied to extract the condensate itself. The way out was to consider
SQCD,evaluate the exact superpotential and then derive the gluino condensate 
using the Konishi anomaly. One more approach concerns the compactification
of one coordinate, find the BPS configurations with two gluino zero modes and 
saturate the condensate by zero modes on these  configurations \cite{khoze}. The different
ways of evaluation of the gluino condensate differ by the numerical factor which
certainly shows that this issue is not understood properly.

The explicit Nekrasov-like evaluation of the instanton sums in the different
dimensions  \cite{nekrasov} allows to attack the issue of the microscopic mechanism for
condensate formation in SUSY YM theory with the new tool. The low energy 
effective action depends on the masses of the matter fields as parameters 
hence the instanton contribution to condensates can be extracted upon differentiation. 
On the other hand using the exact results concerning  equivariant 
K-theory  of  Hilbert scheme of centered points in $\mathbb{C}^2$ \cite{haiman} 
it was found that the torus knot superpolynomials can be represented 
along this way  \cite{gor10,gorneg}. Since K-theory  of Hilbert scheme of points in $\mathbb{C}^2$
is intimately related to the instantons in 5D SYM theory it is natural to assume that
torus knot homologies and  invariants are relevant for some physical observable
in $\Omega$-deformed 5d gauge theory.
In \cite{gm} the instanton-torus knot duality was formulated 
in 5d SQCD based on the
observation made in \cite{bgn}.  It turned out that refined torus knot invariants are involved 
into the formation of the massless flavor condensate.

The  essentially new  findings  in \cite{gm} are as
follows 
\begin{itemize}

\item

It was shown  that the $T_{n,nk+1}$  torus knot superpolynomials
are encoded in the UV properties of the condensate of the massless flavor in
the 5d SYM theory with one compact dimension  and 5d CS term at level k. It
was the first explicit example of the evaluation of the refined knot invariant
in the dual " magnetic" approach in the instanton ensemble  in the gauge theory.

\item

In the previous studies the invariants of the particular knots are involved into
evaluation of the Wilson loops or partition functions and no any summation
over the knot types was needed. In our case due to the instanton-torus knot
duality the summation over instantons implies the summation over  all types
of the torus knots.

\item

In \cite{gm} we travelled across the bridge between the theories with Landau pole at
$N_f=3$ and the asymptotically free theory at $N_f=1$
using the decoupling of heavy degrees of freedom. The heavy flavor in the 
theory with Landau pole can be substituted by the particular observable
which on the other hand can be interpreted as the brane-antibrane pair.
\end{itemize}

In this paper we shall clarify the origin of the instanton-torus knot duality and the role of the
torus knots invariants in the condensate formation. We shall argue that the invariants of the
torus knots provide the entropic factor counting the degeneracies of the particular BPS states. 
Remind that the  pattern for the evaluation of the knot invariants as  counting of BPS states 
has been suggested in \cite{ov}. The useful tool to recognize the knot invariants 
in the topological string framework is the geometric transition \cite{gopa} which occurs
when the set of N topological branes wrapped the submanifold in 3 CY M in $T^*M$ 
gets substituted  for $M= S^3$ by the
resolved conifold with the complex Kahler parameter equals to $Ng_s$
where $g_s$ is the string coupling.  To obtain the knot it is necessary \cite{ov} to add the
Lagrangian brane $L_K$ with the geometry of $S^1 \times R^2$  intersecting the $S^3$ along 
the knot K. Upon the geometric transition the Lagrangian brane remains hence we get
the open A-model setup with $L_K$. It was argued that the HOMFLY polynomials 
count the BPS particles represented by  M2 branes ending on $L_K$.More recently the different aspects 
of the representation of the torus knots in the
topological string framework have been discussed in \cite{marino,klemm,vafatorus}.

Our picture is somewhat close to this approach. We shall demonstrate that HOMFLY polynomials
of the $T_{n,m}$  torus knots in the fundamental representation  count the multiplicity of states 
in the instanton-W-boson web with the
fixed instanton and electric quantum charges $(n,m)$. We consider the 5D SQCD  with the
matter in fundamental and antifundamental representations and the mass of antifundamental
provides the parameter $a=q^N$ in the HOMFLY polynomial \cite{gm}. In our approach the counting 
involves   the enumeration of differently  oriented M2 branes corresponding to  $n$ instantons and 
 $m$ W-bosons. Since the rank of the gauge group in CS representation of 
HOMFLY is represented by mass of the antifundamental  
we shall make the inverse geometric transition representing the Kahler class of the 
blow-up point  in resolved conifold corresponding to the antifundamental matter \cite{vafaold,katz}.
During this inverse geometric transition we substitute the $S^2$ 
 by   the set of topological branes on $S^3$ and
the instanton and W-boson  M2 branes yield  the torus knots in $S^3$.

We  consider the generic electric and instanton quantum numbers and to make 
the instanton particles more tractable
 embed SQED  into SU(2) 5d SQCD 
 when the instantons and W-bosons enter the central charge at the equal footing. The generic picture for arbitrary
quantum numbers can be  also visualized in IIB string theory in terms of the
string webs \cite{kol, sen, verlinde} with boundaries at  the 5-brane web. The instantons and W-bosons
are represented by  different strings obeying the particular rules
of intersection
while the hypers are represented by the combination of the strings and strips.
We should count the number of the different webs with the fixed 
boundary conditions which is equivalent to the counting of the particles
with the different spin and flavor content. The 5-brane web itself can be represented 
by the CY manifold with the degenerated 4-cycle where the edges of the web 
corresponds to the degeneration loci \cite{leung}.
The toric diagram behind the SU(2) gauge
theory with fundamental matter provides the useful insight at the
$n \leftrightarrow m$ duality in the torus knot. Indeed, the diagram is quite symmetric
and the W-bosons gets interchanged with the instantons upon the 90 degree
rotation of the toric diagram. This rotation has to be supplemented by the
change of parameters which has been found in \cite{taki}. This picture explains
the duality between the electric and instantonic quantum numbers in the knot.

We consider the condensate of the fermions from the massless hypermultiplet in 
fundamental $\ll \tilde{\psi}\psi \rr$ in SQED or SQCD. The condensate is electrically neutral 
but  involves the electrically charged
degrees of freedom therefore  can be represented in terms of the Wilson loops in
the first quantized picture. The holomorphy implies that the condensate
is evaluated in the instanton ensemble and  it is
expected to be saturated by the zero modes at the non-perturbative BPS states . Our analysis shows that
the treatment of zero modes requires some care and at fixed value of 
electric and instanton charges there is nontrivial entropic  factor counting the states with
the different spin and flavor content.

It is worth to clarify the place of our approach in the whole subject of
derivation of the knot invariants and homologies within gauge theory. The old
derivation of the Jones polynomial of the knot concerns the evaluation of the
electric Wilson loop in the non-Abelian 3d CS theory \cite{wittenold}. The knot
can be thought of as the trajectory of the particle in some representation $R$ of
gauge group $SU(N)$. The HOMFLY polynomials  colored by the representation $R$
which  are the generating functions for the vev of the Wilson loops  can be
derived in this way. The HOMFLY polynomials can be generalized  to the
superpolynomials introduced in \cite{dgr} which have a clear
interpretation as counting the particular BPS states in the context of the
topological strings \cite{gsv}.The CS  approach has been generalized to the
superpolynomials of the torus knots in \cite{agash} using the matrix model
technique. However the proper field theory yielding the  refined  CS  has
not been found yet.

The alternative "magnetic" S-dual approach has been suggested in
\cite{wittennew,wittengai} where the 4d and 5d SUSY gauge theories provide the
playground for the evaluation of the knot invariants and homologies. It was
assumed in \cite{wittennew,wittengai} that the knot invariants count the
instantons with the particular weights
and the knot itself to some extend corresponds to the magnetic 't Hooft loop of the particular monopole. 
The type of the knot is encoded
in the boundary conditions at the 3d manifold and the differentials in the Khovanov homologies are related to the
multiplicities of the corresponding domain walls.
Our approach formulated in \cite{gm} belongs to the "magnetic"-type  picture. However our theory was
identified as 5D $\suno$ SQED or SQCD with CS term while the theory in
\cite{wittennew,wittengai} was the topologically  twisted $\sunf$ SYM theory.

Let us emphasize that through the paper we shall use the term condensate for the derivative 
of the instanton partition sum $\frac{dZ}{dm}$ with respect to the mass of the hypermultiplet
in the fundamental representation.  Since we are working with the IR effective theory, this term should be taken with some care because
it yields the  fermionic  condensate only in UV. We consider the IR physics 
hence one could have in mind a possible  "contact terms"  which could appear when we  flow from UV to IR.

Completing the Introduction let us present the simplified physical picture behind our calculations.
Although it captures not all ingredients we think that it could be useful for the reader. Consider 
one-loop effective action in the QED in  constant  external electric and magnetic fields. It is just
the fermionic loop in the external field. In a self-dual background field the effective action can 
be identified as the topological string at $T^*S^3$ or equivalently  $SU(N)$ CS at $S^3$ when the rank 
of the  group appears to be the ratio of the fermionic mass and the external field $N \propto \frac{m^2}{eE}$
\cite{gl}.
This is the toy example of the inverse geometrical transition we shall use later and now we have CS with the 
mass dependent rank of the group  inside CY geometry.

\begin{figure}[h]
\centering
\includegraphics[scale=0.9]{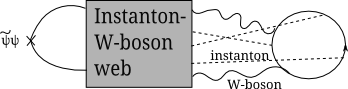}
\caption{Calculating $\ll \tilde \psi \psi \rr$ }
\label{loops}
\end{figure}

Assume now that we have the second fermionic loop in the same external field probably of the
different fermionic flavor. We take the derivative of the second loop with respect to the mass 
which corresponds just to the insertion of  fermionic bilinear(Fig. \ref{loops}).

At the next step we assume that there is the web of interacting particles   of two types 
between the operator insertion and the first loop 
which can braid providing the torus knots $T_{n,m}$ if we have $n$ propagating
particles of one type and $m$ propagating particles of another type. Since we have prepared 
CS theory in CY space from the first loop in the external fields the ends of the propagating 
particles picture the torus knot in $S^3$ inside CY. From the viewpoint of the second loop with
the operator inserted we evaluate the contribution to the condensate from the "tadpole" connected
to the loop by some web involving particles of two types. The knot invariants count the entropy 
of the web 
with fixed two quantum numbers which are attached to the loop in the external field. Equivalently,
it can be thought as the particular entropic factor in the condensate of the bilinear operator.

In our case we have the loop of the antifundamental in the external graviphoton field and the
loop of the fundamental with the inserted bilinear operator due to the derivative in the same external
field. Due to the inverse geometric transition the loop of antifundamental provides the $SU(N)$ CS action 
in CY when the rank of the group is $n \propto \frac{m_a}{\epsilon}$ which is counterpart of the
QED case above. The insertion of the fermionic bililear and the loop of antifundamental 
are connected by the instanton-W-boson web
with electric and instanton charges $(n,m)$ which pictures the torus knot at the
antifundamental side. From the viewpoint of the fundamental we evaluate the condensate of
the bilinear in the external field taking into account the tadpole of the antifundamental connected
by the W-boson-instanton web. The configuration of the web has some peculiarities, for instance,
one has to have in mind that instantons are almost sitting at the top of each other in $C^2$.
The multiplicity of the web yields the entropic factor in the condensate.

One more inspiration from the non-SUSY case goes as follows.
 Remind that the  effective action  QCD yielding the condensate 
in the first quantized representation as the weighted sum of the vev of Wilson loops over the 
arbitrary contours $C$
\beq
\frac{dS_{eff}}{dm}= \frac{d}{dm} \sum_{\mathcal{C}} e^{-mL(C)} e^{i\Phi(C)} \ll W(\mathcal{C}) \rr =\ll Tr \frac{1}{(D-m)} \rr
\eeq
where the electric Wilson loops or the resolvent of the Dirac operator are evaluated in the instanton-anti-instanton
ensemble where  $L(C)$ is the length of the trajectory and $\Phi(C)$  is the spin factor. When we consider the  loop connected by the web with the local operator we can use 
the first quantized picture for  loop hence effectively in this case one evaluates
the weighted correlators of the Wilson loops with the local operator  averaged over the shape of loop and 
over the moduli space of the web. In our case we could have in mind similar representation
in terms of the sum over the averaged  correlators of  supersymmetric Wilson loops with local operator.
.

The paper is organized as follows. In Section 2 we review the duality between the
instantons and the torus knots found in \cite{gm} for the superpolynomials 
for the $T_{n,nk+1}$ series of the torus knots. In Section 3 we summarize the different
ways to get the HOMFLY polynomials for the generic $T_{n,m}$ knots. In Section 4 
we explain how the knot invariants can be obtained from SU(2) SQCD and clarify the
meaning of the $(n,m)$ quantum numbers of the knot as the electric and instanton 
charges. In Section 5 we argue that the condensate can be obtained
from the 5d theory with  the fractional coefficient $k=m/n$  in front of the
5d CS term. Different counting problems yielding the knot polynomials are compared
in Section 6. Also in this section we will discuss various interpretations of knot polynomials. The findings of this paper and the open questions are presented
in the Conclusion.

\section{Instanton - torus knot duality}

In this Section we summarize the key observations from \cite{gm}.
Five-dimensional supersymmetric QED consists of vector field $A_A$, four-component Dirac spinor $\lambda$ and Higgs field $\phi$, all lying in the adjoint representation of $U(1)$.
The Lagrangian reads as follows:
\beq
\mathcal{L}= -\frac{1}{4g^2} F_{AB} F^{AB} + \frac{1}{g^2}(\pr_A \phi)^2 + \cfrac{1}{g^2} \bar \lambda \gamma^A \pr_A \lambda  
\eeq
$\gamma^A, A=1,...,5$ are five-dimensional gamma matrices. Since the adjoint action for the $U(1)$ group is trivial, this is a free theory. 

To introduce $\Omega$-background one can consider a nontrivial fibration of $\mathbb R^4$ over a torus $T^2$ \cite{nekrasov},\cite{NekOk}. The six-dimensional metric is:

\begin{equation}
    ds^2=2dzd\bar z+\left( dx^m+\Omega^md\bar z+\bar\Omega^mdz \right)^2,
    \label{metric_Omega}
\end{equation}
where $(z,\bar z)$ are the complex coordinates on the torus and the four-dimensional vector $\Omega^m$ is defined as:

\begin{equation}
    \Omega^m=\Omega^{mn}x_n,\qquad \Omega^{mn}=\frac{1}{2\sqrt{2}}\begin{pmatrix}0&i\epsilon_1&0&0\\-i\epsilon_1&0&0&0\\0&0&0&-i\epsilon_2\\0&0&i\epsilon_2&0\end{pmatrix}.
     \label{omega}
\end{equation}

In general if $\Omega^{mn}$ is not (anti-)self-dual the supersymmetry in the deformed theory is broken. However one can insert R-symmetry Wilson loops to restore some supersymmetry \cite{NekOk}:

\begin{equation}
    A^I_J=-\frac12\Omega_{mn}\left( \bar\sigma^{mn} \right)^I_J d\bar z-\frac12\bar\Omega_{mn}\left( \bar\sigma^{mn} \right)^I_J dz.
    \label{Wilson}
\end{equation}

The most compact way to write down the supersymmetry transformations and the Lagrangian for the $\Omega$-deformed theory is to introduce 'long' scalars
(do not confuse them with $\mathcal{N}=1$ superfields):

\begin{equation}
    \Phi=\varphi+i\Omega^mD_m, \qquad \bar\Phi=\bar\varphi+i\bar\Omega^mD^m,
      \label{phi_def}
    \end{equation}

We can couple this theory to fundamental hypermultiplet, which consists of two scalars $Q$, $\tilde Q$ and two Weyl fermions $\psi$ and $\tilde \psi$ and 
characterized by two masses: $m$ and $\tilde m$, since $\mathcal{N}=2$ hypermultiplet is build from two $\mathcal{N}=1$ hypermultiplets with opposite
charges. Now the bosonic part reads as:
\beq
\begin{split}
\mathcal{L}_m=-\frac{1}{4g^2}F_{mn}F^{mn}+\cfrac{1}{g^2}(\pr_m \phi + F_{mn} \Omega^n)(\pr^m \phi - F^{mn} \Omega^n)+ \\
   \cfrac{1}{2} |D_m Q|^2 + \cfrac{1}{2} |D_m \tilde Q|^2 + \cfrac{2}{g^2}(i \pr_m(\Omega^m \bar \phi+ \Omega^m \phi)+g^2(\bar Q Q-\bar \tilde Q \tilde Q) )^2 + \\
 \frac{1}{2}|(\phi-m-i \Omega^m D_m)q|^2 + \frac{1}{2}|(\phi- \tilde m-i \Omega^m D_m)\tilde q|^2 + 2g^2|\tilde q q|^2
\end{split}
\eeq

In what follows we will be interested in the condensate of the massless fundamental $<\psi \tilde {\psi}>$
from the 4D viewpoint which depends on the parameters of the model. Upon 
reduction to the 4d theory we have asymptotically free theory when $N_F\leq 2$ for U(1) theory
and $N_f\leq 4$ for SU(2). We shall consider the different number of flavors which
in some case correspond to the theory with Landau pole.

Since we shall count the BPS states it is necessary to remind the spectrum of BPS particles
in the theory. The corresponding central charge involves the quantum numbers corresponding
to the instantons, W-bosons and fundamentals \cite{seiberg}
\beq
Z= \frac{1}{g^2} n_I + n_e a + \sum_i n_{f_i} m_{f_i}
\eeq
The instantons in 5d theory are particles which carry the charge corresponding to the conserved topological current
\beq
J= * TrF\wedge F
\eeq
If we add to the action the CS term
\beq
S_{CS} =k \int A\wedge F\wedge F
\eeq
it couples the topological charge to the gauge field. In particular, this term implies
that the instanton  particle with  instanton charge $n_I$ carries the electric charge $n_Ik$ hence the central charge
can be written as 
\beq
Z=(n_e +kn_I)a + \frac{1}{g^2}n_ I  + \sum_i n_{f_i} m_{f_i}
\eeq
There are also the dyonic instantons carrying the topological and electric charges which 
are unstable under the blowup  into the tubular  D2 brane. Generically
particle carries  quantum numbers $(n_I, n_e, n_f)$.

In \cite{gm} the new instanton-torus knot duality has been formulated for the
Omega-deformed $\suno$ 5d SUSY QED on $\st$ with the Chern-Simons term at level $k$.
It has been proved that the second derivative of the Nekrasov instanton
partition function with respect to the masses of the hypermultiplets  is the
generating function for the superpolynomials of the torus $T_{n,nk+1}$ knots
where $n$ is the instanton charge.

\beq
\label{super}
\left.\frac{e^{\beta M}}{(1+A)\beta^2}\frac{d^2 Z_{nek}(q,t, m_f,M,m_a,Q,k)}{dM\, dm_f}\right|_{m_f\to 0,
M\to \infty} \medskip \\ = \sum_n Q^n(tq)^{n/2} P_{n,nk+1}(q,t,A)
\eeq
where $m_a,m_f,M$ are masses of three hypermultiplets in antifundamental ($m_a$) and 
fundamental representations ($m_f,M$)
and Q is the counting parameter for the instantons. The mapping 
between the parameters at the lhs and rhs goes as follows

\begin{eqnarray}
t=\exp(-\beta \epsilon_1) \\
q=\exp(-\beta \epsilon_2) \\
A=-\exp(\beta m_a) \\
Q=\exp(-\beta/g^2)
\end{eqnarray}
It is worth to think that the information about the knot invariants is encoded 
in the UV properties of the condensate of the massless flavor since the heavy fundamental sets the UV scale M.

The duality implies that the summation over the instanton charge is translated into  summation
over the particular series of the torus knots parameterized by the single integer - instanton charge. Certainly 
one could expect the double sums over generic torus knot $T_{n,m}$ and we shall demonstrate later that the double sum   over the torus knots corresponds to the summation over the instanton and electric charges
at the gauge theory side. It will be clear that the role of heavy flavor in \cite{gm} was to select
the particular value of the electric charge and instead of  a bit artificial procedure in Abelian theory it is more natural to embed the whole picture in SU(2) SQCD when the instanton and W-bosons enter at the equal footing.

In the unrefined
case we  expect the representation of the HOMFLY invariants in terms of the vev of electric
Wilson loops in 3d CS theory and it is desirable to recognize this viewpoint as well. Saying a bit 
differently the question can be formulated as "Where  knots are located?". The answer 
to this question would explain the CS representation of the HOMFLY invariants. We shall 
present the arguments that the knots are represented by the intersection of M2 branes representing 
the BPS states with several quantum numbers with the branes emerging through the inverse
geometric transition. Another picture is provided by the string web ending at the 5-brane web.
The knot invariants count the spin and flavor content of the instanton-W-boson web. We shall also
explain the origin of the AGT type relation between the torus knot invariants and conformal
blocks in q-Liouville theory observed in \cite{gm}.

It is also instructive to recall \cite{gm,bgn} that the differentiation with respect to the heavy mass 
can be equally  thought of as an insertion of the operator $\exp(-\beta \Phi)$. Operator $\Phi$ is not quite the same as adjoint Higgs field since the later is not annihilated by Omega-deformed supersymmetry:
\beq
Q_\Omega \phi = \Omega^m A_m
\eeq
and we require $\Phi$ to be Q-closed: $Q_\Omega \Phi=0$. 
In appendix \ref{app:br}  we will argue that the proper realization of chiral ring operator in the Omega-deformed theory is a
lump of a brane-antibrane system. In what follows we  substitute operator $\exp(-\beta \Phi)$ by a Lagrangian brane which
will be useful to apply different dualities and geometric transitions.

\section{Summary: condensates versus HOMFLY polynomials}

In the previous paper \cite{gm} we focused at the superpolynomials of the torus
knots which measure the response of the condensate of the massless flavor on
the UV scale introduced by the particular operator in the theory with $N_f=2$
or in the $N_f=3$ with one heavy flavor. To some extend this corresponds to
the evaluation of  instanton contribution to the anomalous dimension of the
bilinear operator. However it is interesting to find the interpretation of the
condensate itself in terms of the knot invariants.  We will consider the degeneration of the
superpolynomials to the HOMFLY depending on two generating parameters in the self-dual 
unrefined case. It turns out that 
there are several ways to recognize the HOMFLY invariants of the torus knots 
in the evaluation of the condensates. They are complimentary and can be 
used to clarify the different aspects of the problem.

Let us summarize the different ways how the uncolored HOMFLY polynomials of the
$T_{n,m}$ torus knots in the fundamental representation can be obtained and what is the meaning of the $(n,m)$ quantum
numbers. 
\begin{itemize}

\item 

We can use the representation of the superpolynomials of $T_{n,nk+1}$ knots in terms of the 5d abelian  $N_f=3$ 
gauge theory with the integer CS term  \cite{gm} and consider the limit of self-dual $\Omega$   background $\ep_1+\ep_2=0$ . In this approach we can describe only $T_{n,nk+1}$ series and have  one counting parameter $n$ which corresponds to the 4d instanton charge. The second electric charge is chosen to be equal nk+1 by hands. 

\item

Another approach is suggested by
the celebrated Jones-Rosso
formula \cite{jr93} for the colored HOMFLY-PT polynomial(see \cite{mm_cut} for a nice review). 
We will argue that this representation corresponds to the $N_f=2$ theory without the
additional operator insertions  but with the fractional CS term. In this
representation we shall obtain the $n$-instanton contribution to the condensate
itself as the HOMFLY invariant of $T_{n,nk+1}$ knot when the fractional 5d CS
term is $l=k+1/n$. Note that the denominator in the CS coupling is equal to the number of instantons. This approach does not allow to get the instanton sums but provide the additional 
framework for the evaluations of the separate terms in the instanton sum.
We will describe this approach in Section \ref{sec:cs}.

\item
As we have mentioned above, instead of the insertion of the particular operator in $N_f=2$ theory we can consider  the
$N_f=2$ SU(2) theory supplemented by the Lagrangian brane with zero framing with some value of FI parameter $z$.
To get the HOMFLY polynomials we make two step procedure. First, we consider the decoupling
limit $1/g^2 \rightarrow \infty$ in SU(2) theory when it effectively decouples into the product of
two U(1) theories and pure 4d instantons decouple. However due to the additional Lagrangian
brane we have the  FI parameter which counts the instantons on the Lagrangian brane.
Considering the derivative of the Nekrasov partition function in this case with respect to mass
and expanding it into the double series $z^m Q^n$ we obtain the HOMFLY polynomials
of the generic $(n,m)$ knots as the coefficients of the expansion. In this case parameter $z$ counts 
the 2d instantons while the parameter Q equals to $\exp(\beta a)$ and counts the number
of W- bosons in the decoupling perturbative limit of SU(2) theory. In this approach we can say
that HOMFLY polynomials provide the entropic factor in the condensate in the sector with
the particular defect.  We will develop this approach in Section \ref{sec:lb}

\item
We can embed the abelian theory under consideration into the SU(2) with $N_f=4$. Two masses
of fundamentals are fixed by parameters of the $\Omega$-deformation one mass tends to zero and
one mass is arbitrary. No Lagrangian branes and CS terms are needed in this framework. If we 
expand the derivative of the partition function into the double 
series $e^{m \beta a} Q^n$ corresponding to the expansion in the electric and 
4d instantonic charges we get the HOMFLY polynomial for the generic torus knots. 
No decoupling of 4d instantons occurs since $Q= \exp(\beta g^{-2})$ is finite. This approach is described in Section \ref{to_su2}. It is this picture which immediately explains the origin of relation
with the q-Liouville conformal blocks via AGT relation observed in \cite{gm}
\end{itemize}

It is worth to make more comments concerning the place of the different knot invariants in the context of the evaluation of the condensates. It is known in QCD that  main phenomena
behind the chiral condensate formation is the collectivization of the individual fermionic zero modes in the instanton-antiinstanton ensemble. There is no possibility to get the exact answers in QCD case and one has to restrict 
himself by the effective approaches like the matrix models or low-energy theorems. 
The localization technique in SUSY QCD provides the tool to describe the collectivization of the zero modes 
in the holomorphic ensemble of interacting instantons  in a rigorous way. 

It is clear that the knot invariants provide the entropic factor to the condensate 
which corresponds to the counting of degeneracy of the instanton--W-boson web  with fixed $(n,m)$ 
quantum numbers. This is the particular realization of the approach to the knot 
homologies suggested in \cite{vafa}.  The complete set of states with two 
quantum numbers can be read off   from the string web 
diagram in the IIB approach to the 5d SUSY theory \cite{kol}.The fixed numbers (n,m) 
correspond to the numbers of the F1 and D1 strings involved into the particular web. However there are many possibilities to get the BPS states with these quantum numbers due to the number of string junctions involved and the boundaries of the web selected.

This general picture can be also realized in the combinatorial description of the torus knot 
invariants \cite{ gor10, ors} when the knot invariants including the superpolynomials 
can be derived from the weighted random walks in the $n\times m$ rectangle above the 
diagonal. The each random path corresponds to the particular fixed point in the 
localization integral over the instanton moduli space. The sum over the random paths 
in the 2d Young diagrams can be mapped into the 3d Young diagrams when each path maps to the 
particular 3d Young diagram corresponding to the fixed point. It would be very 
interesting to identify the fixed points with the particular BPS states with three quantum numbers 
explicitly and we hope to discuss this issue elsewhere.

\section{HOMFLY invariants from SU(2) SQCD}

\subsection{Why SU(2)?}

In \cite{gm} we have shown that  some limits of the torus knot invariants are related to the DOZZ factors in the
q-deformed Liouville theory. This implies via the 5d AGT relation that the evaluation of the knot invariants is related 
to the $\Omega$-deformed 5d SU(2) SQCD. On the other hand the previous analysis was based on the abelian theory
hence the relation between the abelian and nonabelian pictures deserves the explanation. This Section is devoted
to this issue and we will argue that the derivation of the  HOMFLY invariants and  condensate 
of the massless flavor matches in two pictures.

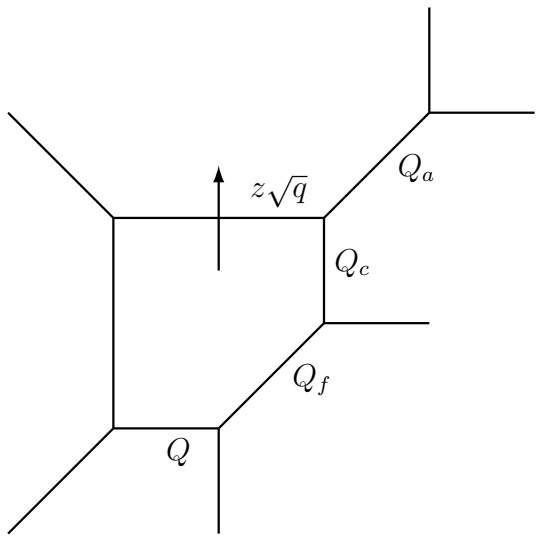
\begin{figure}[h]

\begin{center}

\setlength{\unitlength}{1.4cm}

\begin{picture}(4,5)
\linethickness{0.3mm}
\put(0,1){\line(1,0){1}}
\put(1,0){\line(0,1){1}}
\put(1,1){\line(1,1){1}}
\put(2,2){\line(1,0){1}}
\put(2,2){\line(0,1){1}}
\put(0,3){\line(1,0){2}}
\put(2,3){\line(1,1){1}}
\put(3,4){\line(0,1){1}}
\put(3,4){\line(1,0){1}}

\put(0,1){\line(0,1){2}}

\put(0,1){\line(-1,-1){1}}

\put(0,3){\line(-1,1){1}}

\put(1,2.5){\vector(0,1){1}}

\put(0.5,0.7){\makebox{$Q$}}

\put(2.7,3.4){\makebox{$Q_a$}}

\put(2.1,2.5){\makebox{$Q_c$}}

\put(1.7,1.4){\makebox{$Q_f$}}

\put(1.3,3.2){\makebox{$z \sqrt{q}$}}

\end{picture}

\end{center}

\caption{$SU(2)$ theory with light fundamental and heavy antifundamental hypermultiplets and a Lagrangian brane}

\label{fg:su2}

\end{figure}

To this aim let us remind the toric diagram (aka web of 5-branes in IIB picture) 
for the 5d SU(2) SQCD with some number of flavors. We can obtain the field theory either by
considering this web of branes or by M-theory compactification on the corresponding Calabi-Yau threefold.
These two pictures are related by  "9-11" flip and a chain of T-dualities. 
The diagram is presented at Fig. \ref{fg:su2}  and 
some symmetry corresponding to 90 degree rotation is present supplemented by the particular mapping of parameters.
It the base-fiber duality in the geometrical engineering language \cite{vafa} or the bispectral duality in the
language of the integrable systems. It was discussed in the related framework in \cite{taki,zenkevich} were the 
explicit formulae for the relation between the dual representations of the Nekrasov partition function were derived.
The two Kahler parameters $\exp(-\beta a)$ and $\exp(-\beta/g^2)$ get interchanged under the rotation.
The partition function can be presented as the double sum in the instanton and electric charge numbers and the
sum over instantons gets interchanged with the sum over gauge bosons.

The Lagrangian brane on the internal horizontal line represents M5 brane wrapped around Lagrangian three-cycle in the 
Calabi-Yau: if we consider our toric Calabi-Yau as $T^2 \times \mathbb{R}$ fibration over base $\mathbb{R}^3$, then this Lagrangian cycle is extended  along
$T^2$ and a line in the base $\mathbb{R}^3$ - see \cite{link_vert, disk, hori} for details. 
In IIB language it is represented by the semi-infinite D3 brane perpendicular to the brane-web. From the
5d field theory it looks like the 3d surface defect. Upon the reduction to four dimensions it becomes
familiar semi-infinite D2 brane representing string (see \cite{dgh} for a detailed discussion).

The toric diagram for SU(2) suggests two possible decoupling limits when one or
another K\"ahler class vanishes. These limits correspond to $g^2\rightarrow 0$
and $a\rightarrow \infty$ respectively. In four dimensions one could  say that
two limits correspond to the approaching the perturbative regime. However, the
picture in five dimensions is more involved. We can cut the horizontal line at
the toric diagram (see Fig. \ref{fg:lb} where we showed only one half) which corresponds to the decoupling of the 5d
instanton particles from the partition sum and the  product of two U(1)
partition functions remain. Now the Coulomb modulus in SU(2) theory plays the role of the gauge coupling in the abelian theory and the W-bosons in the
SU(2) theory play now the role of the abelian instantons. Moreover, the Lagrangian brane is placed on the
external leg. 
On the other hand, it was
argued in \cite{dgh} that such decoupling corresponds to decoupling of 5d degrees of freedom
and, therefore, the partition function of the configuration on the Fig. \ref{fg:lb} equals
to the partition function on the 3d defect represented by the surface defect. Therefore, we arrive at 
the kind of 3d/5d duality: 

\begin{center}

5d $\suno$ Abelian theory with $N_f=2$ and Lagrangian brane $\leftrightarrow$ 3d $\sunt$ Abelian theory with $N_f=4$ 
\end{center}

\begin{figure}[h]
\begin{center}
\setlength{\unitlength}{1.4cm}
\begin{picture}(3,5)
\linethickness{0.3mm}
\put(0,1){\line(1,0){2}}
\put(2,1){\line(1,1){1}}
\put(2,1){\line(0,-1){1}}
\put(3,2){\line(0,1){1}}
\put(3,2){\line(1,0){1}}
\put(0,0){\line(0,1){1}}
\put(0,1){\line(-1,1){1}}
\put(1,0.5){\vector(0,1){1}}
\put(0.5,0.5){\makebox{$Q$}}
\put(2.7,1.4){\makebox{$Q_a$}}
\put(1.3,1.2){\makebox{$z \sqrt{q}$}}
\end{picture}
\begin{picture}(3,6)
\linethickness{0.3mm}
\put(2,1){\line(1,0){1}}
\put(2,1){\line(-1,-1){1}}
\put(2,1){\line(0,1){1}}
\put(3,1){\line(1,1){1}}
\put(3,1){\line(0,-1){1}}
\put(4,2){\line(1,0){1}}
\put(4,2){\line(0,1){1}}
\put(2.4,0.6){\makebox{$Q$}}
\put(3.7,1.4){\makebox{$Q_f$}}
\end{picture}
\end{center}
\caption{Two $U(1)$ theories as a limit from $SU(2)$ theory}
\label{fg:u1l}
\end{figure}
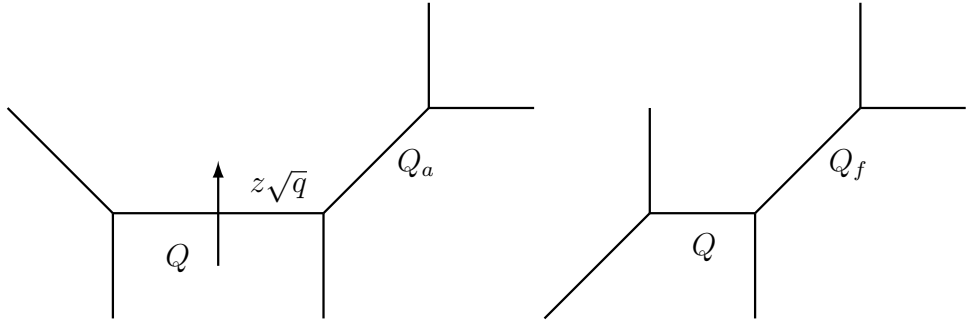

Oppositely one can cut the vertical lines (see Fig \ref{fg:u1l}) and obtain once again the product of two
abelian partition functions where the nonabelain instantons get mapped to the
abelian ones. Note that  the antifundamental matter can be  treated as the
fundamental one in the abelian case. Therefore in the decoupling limit we
find ourself with the product of two abelian theories with some matter content
which depends on the matter content in the initial SU(2) theory.

\begin{figure}[h]

\begin{center}

\setlength{\unitlength}{1.4cm}

\begin{picture}(4,5)

\linethickness{0.3mm}

\put(0,1){\line(1,0){1}}

\put(1,0){\line(0,1){1}}

\put(1,1){\line(2,1){2}}

\put(0,2){\line(1,0){3}}

\put(0,1){\line(0,1){1}}

\put(0,1){\line(-1,-1){1}}

\put(0,2){\line(-1,1){1}}

\put(3,2){\line(3,1){1}}

\put(0.5,0.7){\makebox{$Q$}}

\put(-0.5,1.5){\makebox{$Q_c$}}

\end{picture}

\end{center}

\caption{Pure gauge $SU(2)$ theory with the CS term at level 2}

\label{fg:cs}

\end{figure}
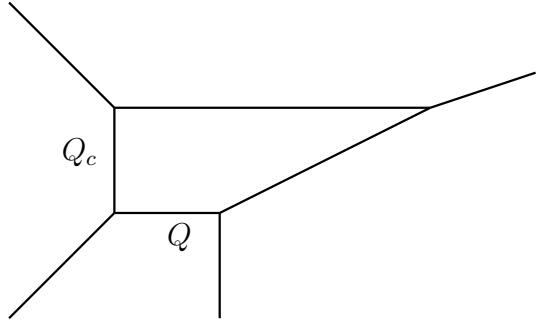

How the decoupling procedure can be applied to the our study? First note that
there are two issues which makes our case a bit more involved. There is
non-vanishing 5d CS term in our Largangian which makes the toric diagram
asymmetric (see Fig. \ref{fg:cs}) . Therefore there is the possibility to make
the naive cut of the horizontal line only which yields the abelian factors with
the different CS terms. Secondly when considering the knot invariants we have
to consider the abelian theory with three flavors one of which plays the role
of the "regulator" . It can be substituted by the operator $\exp(-\beta \phi)$
with the "long " scalar which tells that the decoupling of the heavy flavor is
incomplete in the spirit of the example considered in \cite{rastelli}.

\subsection{Back to abelian theory with Lagrangian brane}

\label{sec:lb}
In terms of the toric diagrams the additional heavy flavor is realized in terms
of the Lagrangian brane attached to the vertical line in the toric diagram. The
field theory interpretation of this Largangian brane deserves the separate
study since its interpretation is different compared to the branes attached to
the horizontal or external leg. To some extend it mimics the single excited
W-boson in the SU(2) corresponding to the  $\exp(-\beta \phi)$ which has such
remnant in the U(1) theory. Hence the decoupling in this toric diagram for
SU(2) theory supplemented by the Lagrangian brane provides the explanation of
the relation between  the torus knot invariants and the vev of the particular
observable in the q-deforned Liouville theory.

The realization of the knot invariants in terms of the SU(2) SQCD supplemented
by defect is useful for the interpretation in terms of the instanton ensemble.
In the abelian case the point-like instanton solution is purely defined and
need for some regularization via non-commutativity or blow-ups at some points.
In SU(2) case it is much simply to think about instantons and the decoupling $a\rightarrow
\infty$ limit means that we effectively are in the perturbative regime with
 instantons and the single electric W-boson excitation. It is this
series of terms in the double sum representation of the Nekrasov partition
function for SU(2) SQCD with $N_f=2$ is intimately related with the torus knot
invariants.

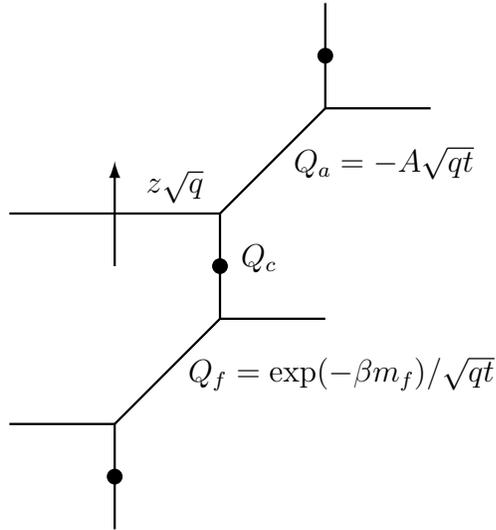
\begin{figure}[h]
\begin{center}
\setlength{\unitlength}{1.4cm}
\begin{picture}(4,5)
\linethickness{0.3mm}
\put(0,1){\line(1,0){1}}
\put(1,0){\line(0,1){1}}
\put(1,1){\line(1,1){1}}
\put(2,2){\line(1,0){1}}
\put(2,2){\line(0,1){1}}
\put(0,3){\line(1,0){2}}
\put(2,3){\line(1,1){1}}
\put(3,4){\line(0,1){1}}
\put(3,4){\line(1,0){1}}
\put(1,2.5){\vector(0,1){1}}
\put(2.7,3.4){\makebox{$Q_a=-A \sqrt{qt}$}}
\put(2.2,2.5){\makebox{$Q_c$}}
\put(1.7,1.4){\makebox{$Q_f=\exp(-\beta m_f)/\sqrt{qt}$}}
\put(1.3,3.2){\makebox{$z \sqrt{q}$}}
\put(2,2.5){\circle*{0.15}}
\put(3,4.5){\circle*{0.15}}
\put(1,0.5){\circle*{0.15}}
\end{picture}
\end{center}
\caption{$SU(2)$ theory with a Lagrangian brane with zero framing after sending $1/g^2 \ra \infty$. Dot indicates preferred direction.}
\label{fg:lb}
\end{figure}

Now let us compute the condensate of the massless hypermultiplet in case of perturbative $SU(2)$ theory supplemented with the Lagrangian brane - Fig. \ref{fg:lb}.

The full partition function reads as:(see \ref{app:vert} for a very brief introduction to the topological vertex):
\beq
Z=\sum_{\la \mu \nu \alpha} (-Q_c)^{|\la|} (-Q_a)^{|\alpha|} (-Q_m)^{|\mu|} C_{\nu \mu \la}(t,q) C_{\emptyset \mu^t \emptyset }(q,t) C_{\emptyset \alpha \lambda^t}(q,t)
C_{\emptyset \alpha^t \emptyset}(t,q) s_\nu(-z \sqrt{q}) 
\eeq
It is convenient to normalize the partition function:
\beq
Z_{inst}=\cfrac{Z}{Z(Q=0)}
\eeq
Then the condensate has the following expansion:
\beq
\ll \tilde{\psi} \psi \rr_{LB}=\cfrac{\pr Z_{inst}}{\pr m_f} \biggl{\vert}_{m_f=0}= \sum_{n,m} Q_c^n z^m P_{n,nk+m}(A,q,t)
\eeq

\begin{eqnarray}
\label{new}
P(A,q,t)_{n,nk+m}= \\ \nonumber
 \mathlarger{ \sum_{\la : |\la|=n }  \cfrac{t^{(k+1) \sum l} q^{(k+1) \sum a}(1-t)\prod^{0,0}(1+A q^{-a'}t^{-l'})\prod^{0,0}(1-q^{a'}t^{l'})}{\prod(q^a-t^{l+1})\prod(t^l-q^{a+1})} \times } \\ \nonumber 
\mathlarger{ Coef_{z^m} M(z) } 
\end{eqnarray}
where $M(z)$ is the contribution from the Lagrangian brane with zero framing:
\beq
M(z) = \prod_{j=1}^{l(\la)} \cfrac{1-z t^{j-1} q^{\la_j}}{1-z t^{j-1}} 
\eeq

This expression has remarkable properties:

\begin{itemize}

 \item At $m=1$ we recover the previous formula for a superpolynomial for $(n,nk+1)$ torus knot.

  \item It gives a polynomial in $A,q,t$ with integer positive coefficients if $gcd(n,nk+m)=1$. Unfortunately, we can
not prove this statement rigorously.

  \item At $k=0$, $P_{n,m}=P_{m,n}$.

  \item At $t=1/q$ it gives correct HOMFLY polynomial for $(n,nk+m)$ torus knot. We will prove this fact in the 
  appendix \ref{app:b}

  \item \textit{However, in general, it does not reproduce conventional superpolynomial.} Another problem is that the representation
in terms of three numbers $n,k,m$ is redundant. Again, for general $\Om$-deformation we will obtain different answers 
for the same knot if we choose $n,k$ and $m$ differently. Nonetheless, in the unrefined case the resulting expression does not
depend on the choice of $n,k,m$.

\end{itemize}

Also, let us note that we could have placed the Lagrangian brane on the leg between $Q_c$ and $Q_f$. In this case
we would get $\tilde M(z)$ instead of $M(z)$:
\beq
\tilde M(z) = \prod_{i=1}^{l(\lambda^t)} \cfrac{1-z q^{i-1} t^{\lambda^t_i}}
{1-z q^{i-1}}
\eeq
It is easy to check that it will again lead to the HOMFLY polynomial, but in different normalization, however.

In fact, explicit formula for the superpolynomial of $(n,m)$ torus knot is known in mathematical literature\cite{gorneg}:

\begin{eqnarray}
P^{(n,m)}=\sum_{|\la|=n} \cfrac{q^{2\sum a} t^{2\sum l} \prod^{0,0}(1+A q^{-a'} t^{-l'})(1-q^{a'} t^{l'})  }
{\prod(q^{a+1}-t^l)(t^{l+1}-q^a)} \times \\ \nonumber 
\sum^{SYT}_{of shape \ \la} \cfrac{\prod_{i=1}^n \chi_i^{S_{m/n}(i)}(1-q t \chi_i) }
{\prod_{i=1}^{n-1}(1-\chi_i)(1-qt \frac{\chi_2}{\chi_1})...(1-qt \frac{\chi_n}{\chi_{n-1}})} \prod_{1 \leq i < j \leq n}
\cfrac{(\chi_j - q \chi_i)(\chi_j-t \chi_i)}{(\chi_j-\chi_i)(\chi_j- q t \chi_i)}
\end{eqnarray}
where
\beq
S_{m/n}(i) = \floor{\cfrac{i m }{n}} - \floor{\cfrac{(i-1) m}{n}}
\eeq
The second factor is a sum over standard Young tableaux of shape $\la$: each tableaux is a Young diagram 
where each box is assigned a number from $1$ to $n$ in such a way that if we travel upwards or rightwards the numbers
decrease. For each $i$ there is a box $\square_i$ and $\chi_i$ equals to $q^{a'} t^{l'}$. 

We see that this quite complicated factor 
corresponds to the contribution of the defect responsible for superpolynomial. Unfortunately we can not identify  this defect in the refined case precisely  however in the self-dual $\Omega$-background it degenerates to the conventional Lagrangian brane.

\subsection{Stable limit}
Let us consider the limit of large Chern-Simons coupling $k \ra \infty$. In this regime instantons die out and contributions from fundamental
and antifundamental matter will factorize. 

On the other hand, it was conjectured in \cite{dgr} that if we consider the so-called stable limit $n \ra \infty$ of the superpolynomial $\mathcal{P}_{n,m}$ we will
obtain unknot colored in the symmetric representation $[m]$:
\beq
\label{eq:sdgr}
\lim_{n \ra \infty} \mathcal{P}_{n,m} = \mathcal{P}_{unknot}^{[m]}
\eeq

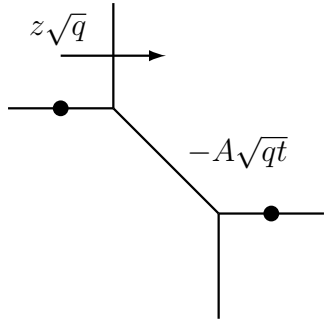
\begin{figure}[h]
\begin{center}
\setlength{\unitlength}{1.4cm}
\begin{picture}(3,3)
\linethickness{0.3mm}
\put(0,2){\line(1,0){1}}
\put(1,2){\line(0,1){1}}
\put(1,2){\line(1,-1){1}}
\put(2,1){\line(1,0){1}}
\put(2,1){\line(0,-1){1}}
\put(0.5,2.5){\vector(1,0){1}}
\put(1.7,1.5){\makebox{$-A \sqrt{q t}$}}
\put(0.2,2.7){\makebox{$z \sqrt{q}$}}
\put(0.5,2){\circle*{0.15}}
\put(2.5,1){\circle*{0.15}}
\end{picture}
\end{center}
\caption{Stable limit. Dot indicates preferred direction.}
\label{fg:stab}
\end{figure}

Now we will show that there is an analogue of this relation in our picture
. Indeed, if we assume $q,t < 1$, only $\la=\varnothing$ gives non-zero contribution in eq. (\ref{new}). 
Fundamental hypermultiplet produces simple
perturbative contribution 
\beq
N_f=\prod_{i=1,j=1}^\infty(1-\exp(-\beta m_f) q^{i-1} t^{-j})
\eeq
which we will discard. Let us look closely at antifundamental hypermultiplet and Lagrangian
brane - see Fig. \ref{fg:stab}. As in the previous sections, we will concentrate on instanton
part of the partition function. However, in this case we will divide by $Z(z=0)$.

This is nothing else than the familiar Ooguri-Vafa geometry which indeed produces colored HOMFLY in the unrefined case\cite{ov}: the partition function
\beq
Z^{unrefined} = \sum_m H_m(A,q) z^m 
\eeq
is the sum of HOMFLY polynomials $H_m(A,q)$ for unknot colored in symmetric representation $[m]$.
Since we have only one 
Lagrangian brane we obtain only symmetric representations. 

In the refined case the situation is a bit more subtle, since the answer depends on the choice of 
preferred direction\cite{ak:change}. Moreover, there
are certain problems with superpolynomial not in the fundamental representation - see \cite{mm_cut} for discussion.
Surprisingly, if we carefully trace the contribution of the brane,  our choice of preferred direction will rather lead us
to superpolynomial in totally \textit{antisymmetric} 
representation $1^m$ \ \footnote{There is no contradiction: in the HOMFLY case there is no difference between
totaly symmetric and antisymmetric representations. Also, for antisymmetric representations the answer does not 
depend on the preferred direction\cite{ak:change}}:
\beq
Z^{refined}= \prod_{i=1}^\infty  \cfrac{ 1-z t^{i-1} }{ 1+A z \sqrt{q} t^{i-1/2} }
\eeq

Therefore, we can write down a relation similar to the eq.(\ref{eq:sdgr}):
\beq
\lim_{k \ra \infty} \cfrac{Z(A,q,t,z,Q)}{Z(A,q,t,z=0,Q)} = \sum_m z^m  P_{unknot}^{1^m}(A,q,t)
\eeq

\subsection{Vortex counting and Lagrangian brane}

As was shown in \cite{dgh} a Lagrangian brane corresponds to a surface operator on the gauge theory side and vortex counting in a 3d theory on this defect matches with the topological vertex computation in the Nekrasov-Shatashvili limit $t \to 1$. So it is instructive to find the expressions for torus knot polynomials directly from vortex counting. In our case we have a three-dimensional $\mathcal{N}=2$ abelian Higgs model on $R^2 \times S^1$ with the addition of one fundamental chiral multiplet that corresponds to 5d vector multiplet and two anti-fundamental ones which comes from 5d fundamental and anti-fundamental hypermultiplets. The parameters for the matter multiplets can be read off from the brane construction of a surface operator in 4d \footnote{Note that in the brane construction the lagranigian brane is replaced to the bottom leg}:

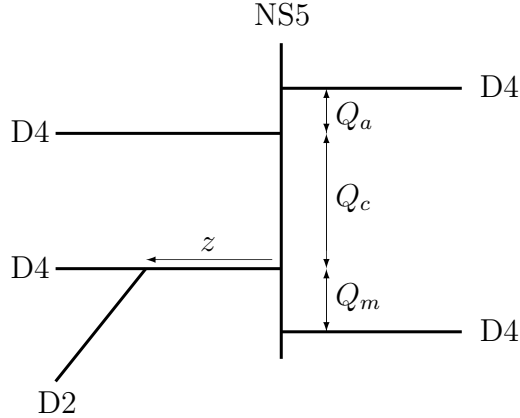
\begin{figure}[h]
\begin{center}
\setlength{\unitlength}{1.2cm}
\begin{picture}(4,5)
\linethickness{0.4mm}
\put(0.5,1.5){\line(1,0){2.5}}
\put(3,0.5){\line(0,1){3.5}}
\put(3,3){\line(-1,0){2.5}}
\put(3,3.5){\line(1,0){2}}
\put(3,0.8){\line(1,0){2}}
\put(1.5,1.5){\line(-0.8,-1){1}}
\linethickness{0.1mm}
\put(3.5,1.7){\vector(0,1){1.3}}
\put(3.5,2.9){\vector(0,-1){1.4}}
\put(3.5, 3){\vector(0,1){0.5}}
\put(3.5, 3.4){\vector(0,-1){0.4}}
\put(3.5, 0.8){\vector(0,1){0.7}}
\put(3.5, 1.4){\vector(0,-1){0.6}}
\put(2.9, 1.6){\vector(-1,0){1.4}}
\put(2.7,4.2){\makebox{NS5}}
\put(5.2 ,3.4){\makebox{D4}}
\put(5.2 ,0.7){\makebox{D4}}
\put(0, 1.4){\makebox{D4}}
\put(0, 2.9){\makebox{D4}}
\put(0.3, -0.1){\makebox{D2}}
\put(2.1,1.7){\makebox{$z$}}
\put(3.6, 2.2){\makebox{$Q_c$}}
\put(3.6, 1.1){\makebox{$Q_m$}}
\put(3.6, 3.1){\makebox{$Q_a$}}
\end{picture}
\end{center}
\caption{The brane construction of a surface operator in type IIA string theory.}
\label{fg:d2surf}
\end{figure}

We also introduce the Omega-background parameter $q=\exp({-\beta \epsilon})$. The result for the vortex partition function is the following:

\beq
Z(z,q; Q_c, Q_m, A) = \sum_{m=0}^{\infty} \frac{\prod_{j=0}^{m-1} (1-Q_m q^j) \prod_{j=0}^{m-1} (1+A Q_c q^j)}{\prod_{j=1}^{m} (1 - q^j)\prod_{j=0}^{m-1} (1-Q_c q^j)} z^m
\eeq
with a shift $Q_c  \to q^{-1} Q_c$. Taking the derivative with respect to $m_f$ and the limit $m_f \to 0$ we obtain 
\beq
\cfrac{\pr Z}{\pr m_f} \biggl{\vert}_{m_f=0}=  \sum_{m=0}^{\infty} \frac{ \prod_{j=0}^{m-1} (1+A Q_c q^j)}{ (1 - q^m)\prod_{j=0}^{m-1} (1-Q_c q^j)} z^m = \sum_{m,n} H_{(m,n)}(A,q) z^m Q_c^n
\eeq
The coefficients $H_{(m,n)}$ can be found easily from the expression above
\beq
H_{(m,n)} = \frac{1+A}{1-q} \sum_{k}  q^{k(k+1)/2} \frac{[m+n-k-1]_{q}!}{[m]_{q} [n]_{q} [k]_{q}! [m-k-1]_{q}! [n-k-1]_{q}!} A^k 
\eeq
where 
\beq
[n]_q = \frac{1-q^n}{1-q}
\eeq
It coincides with the result obtained using topological vertex and reproduces the known expressions for HOMFLY polynomials for coprime $(m, n)$. Also note that if $A=0$ then
\beq
(1-q) H_{(n,n+1)} = \frac{[2 n]_{q}!}{[n]_{q}! [n+1]_{q}!}
\eeq
that is a $q$-deformed Catalan number.

Thus we obtain that knot polynomials also appear in the expansion of the condensate in 3d gauge theory. Now the winding numbers of a torus knot corresponds to the vortex parameter $z$ and the flavor parameter $Q$.

\subsection{From Lagrangian brane to $SU(2)$ theory with four flavours}

\label{to_su2}
According to the AGT conjecture \cite{agt} and its 5-dimensional generalization \cite{awata,aandb,wyllard,taki},
the perturbative part of $SU(2)$ Nekrasov partition function is equal to three-point function 
in the Liouville theory or its q-deformed analogue. What is more, the insertion of a surface defect
corresponds to  the insertion of operator $V_{2,1}$ which is degenerate at level 2 .
So we conclude that the perturbative partition function with a surface defect should be equal to the full $SU(2)$ 
partition function but with a very special choice of fundamental masses. Indeed such an equivalence was conjectured
and checked in \cite{taki_bubb} by a virtue of two  transformations on topological vertexes. Now we are
going to review them.

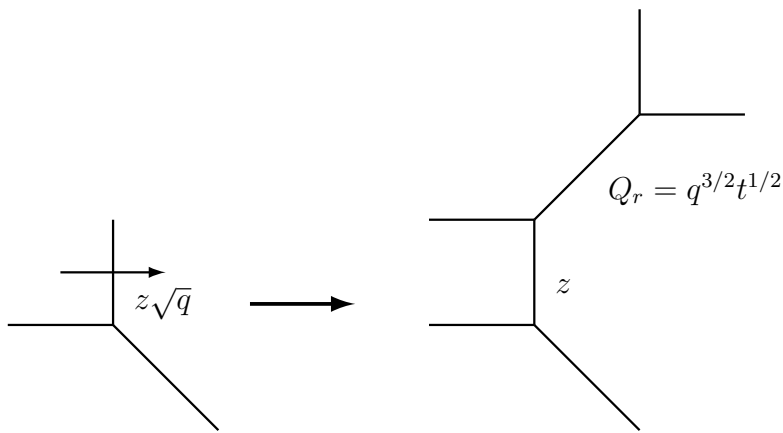
\begin{figure}[h]

\begin{center}

\setlength{\unitlength}{1.4cm}

\begin{picture}(8,4)

\linethickness{0.3mm}

\put(0,1){\line(1,0){1}}

\put(1,1){\line(0,1){1}}

\put(1,1){\line(1,-1){1}}

\put(0.5,1.5){\vector(1,0){1}}

\linethickness{0.5mm}

\put(2.3,1.2){\vector(1,0){1}}

\linethickness{0.3mm}

\put(4,1){\line(1,0){1}}

\put(5,1){\line(0,1){1}}

\put(5,1){\line(1,-1){1}}

\put(4,2){\line(1,0){1}}

\put(5,2){\line(1,1){1}}

\put(6,3){\line(1,0){1}}

\put(6,3){\line(0,1){1}}

\put(1.2,1.15){\makebox{$z \sqrt{q}$}}

\put(5.7,2.2){\makebox{$Q_r=q^{3/2} t^{1/2}$}}

\put(5.2,1.3){\makebox{$z$}}

\end{picture}

\end{center}

\caption{Refind open-closed duality}

\label{fg:op_cl}

\end{figure}

The first one is usual open-closed duality: we can substitute a Lagrangian brane by a resolved conifold - see Figure \ref{fg:op_cl}.

Actually, if we consider general $Q_r$ we will arrive at the following contribution(the second ratio is a normalization
by a perturbation contribution):

\beq
\prod_{i=1}^{l(\la)} \prod_{j=1}^{+\infty} \cfrac{(1-z Q_r t^{i-3/2} q^{\la_i-j-1/2})}{(1-z t^{i-1} q^{\la_i-j})}
\cfrac{(1-z t^{i-1} q^{-j})}{(1-z Q_r t^{i-3/2} q^{-j-1/2})}
\eeq
If we take $Q_r=q^{3/2} t^{1/2}$ we obtain exactly $M(z)$.

\begin{figure}[h]

\begin{center}

\setlength{\unitlength}{1.4cm}

\begin{picture}(8,4)

\linethickness{0.3mm}

\put(0,0){\line(1,1){1}}

\put(1,1){\line(1,0){1}}

\put(2,1){\line(1,-1){1}}

\put(1,1){\line(0,1){1}}

\put(1,2){\line(-1,1){1}}

\put(2,1){\line(0,1){0.5}}

\put(1,2){\line(1,0){0.5}}

\linethickness{0.5mm}

\put(3,1.2){\vector(1,0){1}}

\linethickness{0.3mm}

\put(4,0){\line(1,1){1}}

\put(5,1){\line(1,0){1}}

\put(6,1){\line(1,-1){1}}

\put(5,1){\line(0,1){1}}

\put(5,2){\line(-1,1){1}}

\put(6,1){\line(0,1){1}}

\put(5,2){\line(1,0){1}}

\put(6,2){\line(1,1){1}}

\put(7,3){\line(1,0){1}}

\put(7,3){\line(0,1){1}}

\put(6.4,2.2){\makebox{$q^{1/2} t^{1/2}$}}

\end{picture}

\end{center}

\caption{Projector onto trivial representations}

\label{fg:proj}

\end{figure}
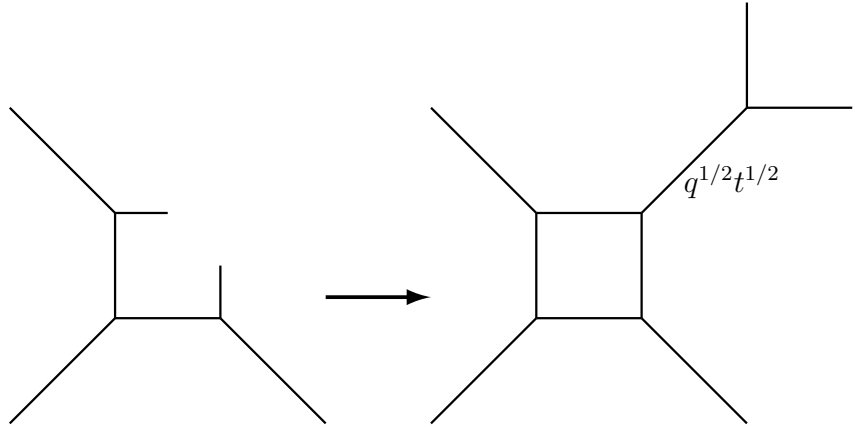

The second transformation is a construction of the projector onto trivial representations: suppose
we have a diagram where two external lines intersect. Then we can add additional resolved conifold with
a special K\"ahler class which will project representations on these two lines onto trivial ones - see Figure \ref{fg:proj}

To sum up, we can obtain our "almost superpolynomial" simply by $SU(2)$ theory with four flavours.

\begin{figure}[!h]

\begin{center}

\setlength{\unitlength}{1.2cm}

\begin{picture}(7,6)

\linethickness{0.3mm}

\put(2,2){\line(1,0){1}}

\put(3,1){\line(0,1){1}}

\put(3,2){\line(1,1){1}}

\put(4,3){\line(1,0){1}}

\put(4,3){\line(0,1){1}}

\put(2,4){\line(1,0){2}}

\put(4,4){\line(1,1){1}}

\put(5,5){\line(0,1){1}}

\put(5,5){\line(1,0){1}}

\put(2,2){\line(0,1){2}}

\put(2,2){\line(-1,-1){1}}

\put(2,4){\line(-1,1){1}}

\put(1,1){\line(-1,0){1}}

\put(1,1){\line(0,-1){1}}

\put(1,5){\line(-1,0){1}}

\put(1,5){\line(0,1){1}}

\put(2.5,1.7){\makebox{$Q$}}

\put(4.7,4.4){\makebox{$Q_a$}}

\put(4.1,3.5){\makebox{$Q_c$}}

\put(3.7,2.4){\makebox{$Q_f$}}

\put(3,4.2){\makebox{$z$}}

\put(1.6,1.2){\makebox{$\sqrt{q t}$}}

\put(0.7,4.2){\makebox{$q \sqrt{q t}$}}

\end{picture}

\end{center}

\caption{$SU(2)$ theory with four flavours}

\label{fg:su2_m4}

\end{figure}
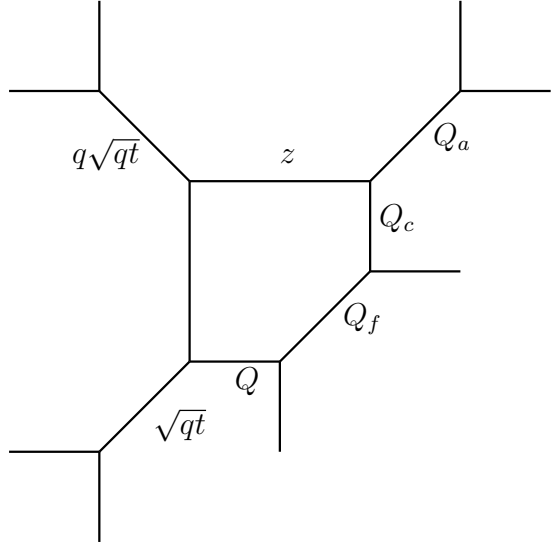
\begin{figure}[!h]

\begin{center}

\setlength{\unitlength}{1.8cm}

\begin{picture}(8,4)

\linethickness{0.3mm}

\put(1,1){\line(1,0){0.7}}

\put(1.7,1){\line(1,1){0.3}}

\put(1.7,1){\line(0,-1){0.5}}

\put(2,1.3){\line(0,1){0.7}}

\put(2,1.3){\line(1,0){0.5}}

\put(2,2){\line(1,1){0.5}}

\put(2.5,2.5){\line(1,0){0.5}}

\put(2.5,2.5){\line(0,1){0.5}}

\put(1,2){\line(1,0){1}}

\linethickness{0.5mm}

\put(1.3,1.7){\vector(0,1){0.6}}

\put(3,1.2){\vector(1,0){1}}

\linethickness{0.3mm}

\put(4.5,1){\line(1,0){1.2}}

\put(5.7,1){\line(1,1){0.3}}

\put(5.7,1){\line(0,-1){0.5}}

\put(6,1.3){\line(0,1){0.7}}

\put(6,1.3){\line(1,0){0.5}}

\put(6,2){\line(1,1){0.5}}

\put(6.5,2.5){\line(1,0){0.5}}

\put(6.5,2.5){\line(0,1){0.5}}

\put(5.3,2){\line(1,0){0.7}}

\put(5.3,2){\line(-1,-1){0.3}}

\put(5.3,2){\line(0,1){0.5}}

\put(5,1.7){\line(0,-1){1.2}}

\put(5,1.7){\line(-1,0){0.5}}

\put(6.3,2.1){\makebox{$Q_m$}}

\put(5.9,1.0){\makebox{$A$}}

\put(4.9,2.0){\makebox{$q^p$}}

\put(4.8,1.1){\makebox{1}}

\end{picture}

\end{center}

\caption{A more general picture in the unrefined limit}
    
\label{fg:stack}

\end{figure}
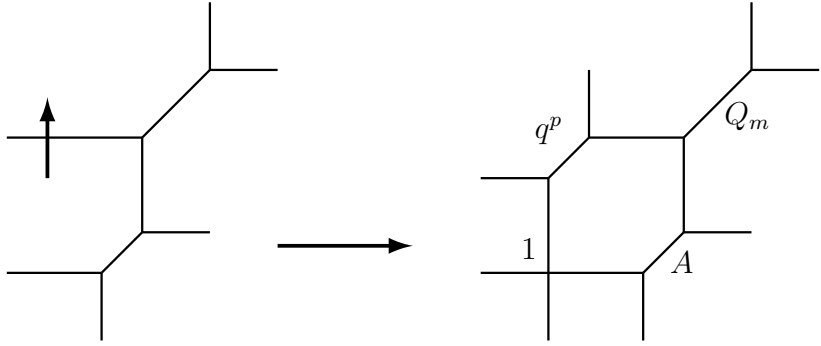

To obtain a more general picture for SU(2) theory we can consider not a single lagrangian brane, but a stack of $p$ branes. For this case we specialize to the unrefined limit. After doing the same procedure we obtain SU(2) theory \cite{dgh} with two arbitrary mass parameters $q^p$ and $A$ (see fig. \ref{fg:stack}). 

Note that we changed the order of matter multiplets on a toric diagram. This change can be considered as a replacement of a Lagrangian brane from a bottom leg to the top leg. It doesn't affect the quantity we compute. In fact we already did this change when we were considering vortex counting.

Again let us compute the derivative of the free energy with respect to $Q_m$. The answer can be expressed in the following way
\beq
\cfrac{\pr Z(1,A,q^p,Q_m; q)}{\pr m} \biggl{\vert}_{m=0}= \sum_{(m,n)}  \, H_{(m,n)}(A,q) H_{(m,n)}(q^p,q) \, Q_B^m Q_F^n 
\eeq
where as above $H_{(m,n)}$ is a HOMFLY polynomial for torus knot. So the stack of branes simply gives another factor in the derivative which is a knot polynomial. This factor is trivial for a single brane.

\begin{figure}[!h]
\begin{center}
\setlength{\unitlength}{1.3cm}
\begin{picture}(5,5)
\linethickness{0.4mm}
\put(1.9,1){\line(1,0){2.5}}
\put(3,0.5){\line(0,1){0.4}}
\put(3,1.1){\line(0,1){1.9}}
\put(3,3){\line(-1,0){2.0}}
\put(0.8,3){\line(-1,0){0.3}}
\put(3,3){\line(1,1){0.5}}
\put(3.5,3.5){\line(1,0){1}}
\put(3.5,3.5){\line(0,1){1}}
\multiput(3,1.4)(-0.16,-0.1){4}{\line(-0.8,-0.5){0.1}}
\multiput(3,1.45)(-0.16,-0.1){5}{\line(-0.8,-0.5){0.1}}
\multiput(1.5,3)(-0.16,-0.1){4}{\line(-0.8,-0.5){0.1}}
\multiput(1.42,3)(-0.16,-0.1){3}{\line(-0.8,-0.5){0.1}}
\put(0.9,2.0){\line(0,1){2}}
\put(2.5,1.4){\makebox{$S^3$}}
\end{picture}
\end{center}
\caption{Resolved conifold with two stacks of lagrangian branes}
\label{fg:stack2}
\end{figure}
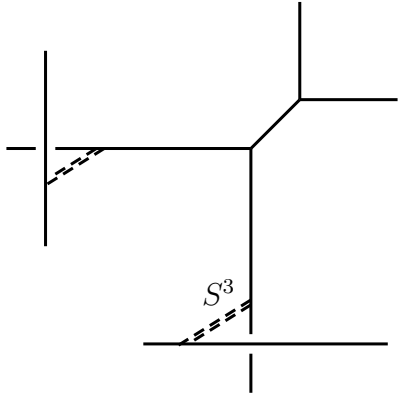

To relate our approach to the conventional CS representation  
we make a reversed  geometric transition  replacing a matter multiplet $A$ by an another stack of branes (see fig. \ref{fg:stack2})

Thus each stack of branes gives us a HOMFLY, which also has an interpretation as a Wilson loop operator for torus knot 
in Chern-Simons theory on $S^3$ which lives on each stack of branes with the parameters that 
coincide with the parameters in our case. The explanation of this fact is given in section \ref{bound}.

Let us comment on the inverse geometrical transition. Geometric engineering of the fundamental matter implies
$N_f$ blow-ups wth the corresponding Kahler moduli fixed by masses of fundamentals and antifundamentals.
We aim to get CS Lagangian therefore we perform the inverse geometric transition and trade the Kahler moduli
into the number of the topological branes wrapped around $S^3$. The relation between parameters goes as follows
\beq
N g_s=m_a
\eeq
where it is useful to perform the inverse transition with respect to the matter in the antifundamental. Having in mind
the relation between the graviphoton field and the string coupling constant we get required parameter $A$ obeying the relation $A=q^N$.
Note that to some extend similar relation can be seen even in the abelian theory in the external constant self-dual
electromagnetic field $F$. The one-loop effective action is related to the large N limit of CS theory as follows
\beq
S_{eff}= \int_{0}^{\infty} ds e^{-sk}  \l \frac{s/2}{sin(s/2)} \r^2 = log Z_{CS}(N\rightarrow \infty, k, S^ 3)
\eeq
where $k= \frac{m^2}{2eF}$ \cite{gl}. We can consider this case as the phenomena of the same nature having in mind
the rank-level duality. However it is important to investigate the inverse transition in more detailes. In particular in the
context of the knot invariants it is interesting to look at the interference of the inverse transitions performed with the several
flavors.

\subsection{Reduction to D=4 theory}
Let us discuss reduction to the four dimensions in the framework of the $SU(2)$ theory.
In D=5 we have the W-boson and instanton particles propagating in the loop. Both
of them correspond to M2 branes wrapped around the base or fiber cycles. As
was shown in \cite{Nek5d} the one-loop contribution amounts to the instanton
series in D=4 theory where the double sum in electric and instanton charges in D=5 gets reduced to  the single sum
in instanton charge in D=4. Since the HOMFLY polynomial measures the degeneracy of the states with
$(n_I,n_e)$ quantum numbers this reduction implies that the partial
resummation of the knot invariants takes place. Indeed, in order to take 4D limit, we send $\beta \ra 0$ and $g \ra 0$, but keeping the Coloumb parameter and
masses finite:
\begin{eqnarray}
\frac{\beta}{g^2} = \frac{1}{g_{4D}^2} = \const \\
a, m_a, \ep =  \const
\end{eqnarray}
The first equation reflects the fact that instantons propagating along the compact dimension become more familiar point-like instantons in four dimensions.
To obtain particular $n-$instanton contribution $c_n$ to the condensate:
\beq
\ll \tilde{\psi} \psi \rr_{4D}= \sum_n e^{-n/g_{4D}^2}  c_n(\ep,m,a)
\eeq
we need to perform the summation over all possible electric charges:
\begin{equation}
c_n(a,m_a,\ep) = \lim_{\beta \ra 0} \sum_{m=0}^\infty e^{-m a \beta} H_{n,m}(q,A)
\label{limit}
\end{equation}
As we send $\beta \ra \infty$ terms with large $m$ become more and more relevant until the sum turn into Laplace-like transform. However, this is not exactly Laplace 
transform since $q=\exp(-\beta \ep)$ and $A=-\exp(\beta m_a)$ depend on $\beta$ and approach 1 and -1 respectively. Because of this, we loose information
about finite powers of $q$ and $A$ and the four-dimensional limit is not invertible.
Nonetheless, it is natural to ask: is anything left from knot invariant?

The answer is straightforward: it is easy to see from the eq. (\ref{limit}) that the large $m$ behavior(stable limit) of
$H_{n,m}$ is encoded into the analytic structure in variable $a$
of 4D instanton contribution $c_n$. Each pole at $-\alpha \ep$ corresponds to the term $q^{\alpha m}$ in HOMFLY polynomial $H_{n,m}$.

For example, let us consider two instantons. $H_{2,m}$ for $m>0$ reads as:
\beq
H_{2,m}=\cfrac{q(A+q-q^{-m}(Aq+1))}{q^2-1}
\eeq
On the other hand, taking the 4D limit:
\beq
\cfrac{m_a+\ep}{2 \ep a}+\frac{m_a-\ep}{2 \ep(\ep-a)}
\eeq
Two poles at $0$ and $\ep$ reflect $q^0$ and $q^{-m}$ terms respectively.
We see that the four-dimensional limit is sensitive only to the large $m$ growth of the HOMFLY polynomial $H_{n,m}$. It means that the second winding
of the knot become condensed. This ``mathematical'' condensation reflects physical condensation: in four-dimensions the fifth component of the vector potential $A_5$ condensates
and joins the Higgs scalar.

\section{Fractional 5d Chern-Simons term}
\label{sec:cs}
In this Section we shall focus at the case of the fractional Chern-Simons term. It can be
also considered as the fractional framing of the torus knot. Let us emphasize that
the fractional CS term provides a bit different picture compared to
the previous Sections and the level of the CS term  is
related in a different way with the type of the knot.
The Jones-Rosso formula can be rewritten as (see Appendix \ref{app:a} for details):

\beq
\label{homfl2}
H^{(n,m)}_\square(A,q)=(-1)^{n-1}\cfrac{1-q^n}{q^n} \sum_{|\la|=n} \mathlarger{  q^{(\frac{m}{n}+1)\sum(l-a)}
\cfrac{\prod^{0,0}(1-q^{l'-a'}) \prod^{0,0}(1+A q^{a'-l'})}{\prod(q^{-l-1}-q^a)(q^{-l}-q^{a+1})} }
\eeq

The later formula strikingly resembles the instanton partition function of 5D $\mathcal{N}=1$ U(1) gauge theory on 
$\mathbb{R}^4_\Omega \times S^1_\beta$ in self-dual Omega deformation $\ep_1=-\ep_2$ with antifundamental
matter of mass $m_a$ and fundamental matter of mass $m_f$ and with the CS term $m/n$  :
\beq
\cfrac{\pr \tilde{Z}^{inst}_n}{\pr m_f} \at_{m_f=0} = (1+A) \beta \sum_{|\la|=n} \mathlarger{  q^{(\frac{m}{n}+1)\sum(l-a)}
\cfrac{\prod^{0,0}(1-q^{l'-a'}) \prod^{0,0}(1+A q^{a'-l'})}{\prod(q^{-l-1}-q^a)(q^{-l}-q^{a+1})} }
\eeq

with
\beq
q=\exp(-\beta \ep_2),\ A=\exp(\beta m_a)
\eeq
We will denote the partition function of this theory by $\tilde Z$.
For unknot $(n,1)$ this formula gives the following result:
\beq
H^{(n,1)}_\square = \cfrac{1}{(1-q) q^{(n-1)/2}}
\eeq

Now let us recall the following expression for the superpolynomial in the fundamental representation of $(n,nk+1)$ torus knot:

\begin{eqnarray}
P(A,q,t)_{nk+1,n}= \\ \nonumber
 \sum_{\la : |\la|=n } \large{ \cfrac{t^{(k+1) \sum l} q^{(k+1) \sum a}(1-t)(1-q)\prod^{0,0}(1+A q^{-a'}t^{-l'})\prod^{0,0}(1-q^{a'}t^{l'})
(\sum q^{a'} t^{l'})}{\prod(q^a-t^{l+1})\prod(t^l-q^{a+1})} }
\end{eqnarray}

This expression can be obtained as an instanton partition function of 5D  $\mathcal{N}=1$
U(1) gauge on $\mathbb{R}^4_\Omega \times S^1_\beta$  in general Omega-background, with the CS term k, 
with 2 fundamental matters and one
anti-fundamental matter. Or, equivalently, with fundamental matter, antifundamental matter and chiral observable:

\begin{equation}
P(A,q,t)_{n,nk+1}= t^{-n/2} q^{-n/2}(1-t)(1-q) \cfrac{1}{1+A} 
\cfrac{\exp(\beta M)}{\beta^2} \cfrac{\pr}{\pr m_f}\cfrac{\pr}{\pr M}  Z^{U(1)}_n(m_f,m_a,M) \biggl{\vert}_{m_f \ra 0,\ M \ra \infty} 
\end{equation}

or equivalently:
\begin{equation}
P(A,q,t)_{n,nk+1}= t^{-n/2} q^{-n/2} \cfrac{1}{1+A} 
 \cfrac{\pr}{\pr( \beta m_f)} \ll \exp(-\beta \phi) \rr,\ m_f = 0
\end{equation}
We will denote the partition function of this theory by $Z$ without tilde in contrast to the theory with fractional
CS term.

In order to obtain the HOMFLY-PT polynomial one need to take $t=1/q$.
The relation between these two formulas reads as follows:
\beq
P_{n,nk+1}(A,q,q^{-1}) = (-1)^n \cfrac{H^{n,nk+1}_\square(A,q)}{H^{n,1}_\square(A,q)} =(-1)^n (1-q)q^{(n-1)/2} \homfnk  
\eeq
while
\beq
\homfnk = (-1)^{n-1} \cfrac{1-q^n}{(1+A)q^n} \cfrac{\pr \tilde Z_n}{\pr (\beta m_f)} \at_{m_f \ra 0}
\eeq
From the above formulas it is possible to obtain various relations between condensates in different theories.

To complete this Section two remarks are in order. First, one could question about
the generic instanton contribution when the instanton number does not equal to the
denominator of the CS level.  This question has been discussed in the math
literature in \cite{etingof}.  It turns out that the generic situation is quite
complicated and the instanton contributions can be expressed in terms of the
superpositions of the colored HOMFLY polynomials. The second point to be
mentioned concerns some analogy with the FQHE which can be described in two
ways. One involves the fractional 3d CS term while in the second approach when
the composite fermions are introduced the system of new effective degrees of
freedom is described by the integer CS term. It seems that the discussion in
this Section has some common features with that case.

Let us remind that the HOMFLY invariants can be obtained from the viewpoint of
the instanton  quantum mechanics.  This representation corresponds just
to the fractional CS approach. As we have discussed in \cite{gm} the 5d CS term induces
the interaction between the instantons.The interaction is attractive
and its strength is fixed by the coefficient in front of the  CS term $k=m/n$ . In this 
approach the
number of instantons is strictly correlated with the CS term and is equal to $n$. Since the
interaction is attractive  the falling to the center takes place and we have to
investigate the fine structure of the $n$-instantons sitting at one point. The
HOMFLY polynomials correspond to the counting of the $E=0$ states in the Calogero
Hamiltonian \cite{gorsky,gors}.
The special property of the rational CS term is that the corresponding Cherednik algebra has the finite-dimensional 
representation and the Calogero Hamiltonian is expressed in terms of the Dunkl operators $D_i$ which are  generators of  the Cherednik algebra. The HOMFLY polynomials can be considered as the special twisted character of the
finite-dimensional representation which on the other hand is the twisted Willen-like  index in the Calogero model counting 
the $E=0$ states with the proper weights.

\section{Comments on the counting problems}

In our paper we have argued that the HOMFLY polynomials of the torus knots count the
multiplicities of the states with the fixed instanton and electric charges. Let us make a few
remarks concerning  its relation to another counting problems  and
possible applications.

\subsection{Standard picture}

According to \cite{ov} to get the HOMFLY polynomials from the type A topological strings 
one starts with the  $T^*S^3$ geometry with N branes wrapped $S^3$. 
 and add   the Lagrangian brane 
wrapped the Lagrangian manifold $L_K$ intersecting $S^3$  along the knot K. The torus 
knot is the intersection of the singular surface
\beq
x^n=y^m
\eeq
with the $S^3$. The Lagrangian brane has the topology $S^1\times R^2$ in CY manifold
and is identified as the total space to the co-normal bundle to the  knot K. Upon the 
geometric transition the N branes disappear and the resolved conifold supplemented
the Lagrangian brane emerges.
The Lagrangian 
brane lies in the fiber of the resolved conifold.

The HOMFLY polynomial corresponds to the counting of open M2 branes which end 
at the $L_K$ and  have the
topology of disc in the target and can be wrapped around the $P^1$ base. Two generating
parameters count the spin of the open M2 brane and its momentum in the 11-th direction. Due to the singular fiber 
in the case of the torus knot contrary to the unknot calculation in  \cite{ov} the counting
of M2 branes is nontrivial due to the presence of the singularity and to some extend the 
single M2 brane acquires multiplicity. It was shown in \cite{dsv} that this brane picture 
reproduces the approach in \cite{ors}. From the M-theory viewpoint the 
HOMFLY polynomial counts the M2 states ending on the M5 brane with geometry
$R^{2,1}\times L_K$ where $R^{2,1} \in R^{4,1}$. Let us emphasize that in this
conventional approach the knot K is fixed by the Lagrangian M5 brane added by hands.

\subsection{Knots as boundaries of holomorphic instantons}
\label{bound}
In our picture the whole geometry contains the information about all torus knots and each knot corresponds to a particular sector of BPS spectrum. The knot is selected not by the additional $L_K$ brane but by BPS state itself. As we have
discussed above the key point is the inverse geometric transition when we substitute  the blow-up  $S^2$ whose
K\"ahler modulus is fixed by the mass of the antifundamental by the set of topological branes wrapped $S^3$ whose number
is fixed by the mass. After all we count the multiplicities of M2 branes corresponding to the torus knot ending
at the stack of topological branes. The representation of coloring  of the knot corresponds to the way how the M2 brane
ends at the stack. Note that we  try to decouple the antifundamental and send its mass to infinity the
rank of the gauge group tends to infinity as well. Hence even upon the naive decoupling of the heavy flavor
we keep the information about the knot invariants. Note some similarity with the bootstrapping of the heavy
flavor in \cite{rastelli} when the non-Abelian string was the remnant after the decoupling of the heavy flavor.

So it is tempting to relate each knot with the geometry of holomorphic instantons in the corresponding sector. As in \cite{ov} knots are seen in the picture where all branes are involved. So let's consider a resolved conifold with two stacks of branes (see fig. \ref{fg:stack2}). The contribution of holomorphic instantons into the partition function according to \cite{ov} is 

\beq
Z_{inst} =\biggl\langle \exp \biggl( \sum_{\beta, s, R_1, R_2} \sum_{n=1}^{\infty} \mathcal{N}^{\beta}_{s, R_1, R_2} \frac{q^{n s}}{n(1-q^n)} Q_{\beta}^{n} \, Tr_{R_1} U^n Tr_{R_2} V^n  \biggr) \biggr\rangle_{CS_1, CS_2}
\eeq
where $U$ and $V$ are the holonomies of a gauge fields in the 
representations $R_1$ and $R_2$ on two different stacks of branes over the boundaries of instantons 
and $Q_\beta$ is the Kahler parameter of a cycle $\beta$. $\mathcal{N}^{\beta}_{s, R_1, R_2}$ are integer coefficients that 
counts the degeneracies of BPS states with a given spin $s$ and which transform in the representations $R_1$ and $R_2$ 
under a $U(p)_i$ symmetries on the branes. Now consider an instanton that warps $m$ times 
around $Q_B$ cycle and $n$ times around $Q_F$ cycle. It also warps $(m,n)$ cycle on a torus in the fiber
of the toric diagram. So it's boundary also warps $(m,n)$ cycle on a torus on a 
Lagrangian branes. In this way we have a knot on $S^3$ and the contribution of this holomorphic 
instantons into our Chern-Simons theory appears with observables associated with a 
corresponding torus knots. Now if we take the derivative with respect to the mass, 
take the limit $m \to 0$ and expand the result over $Q_B$ and $Q_F$ then the coefficient in front of $Q_B^m Q_F^n$ for coprime $(m,n)$  can be expressed in the following way:

\beq
\frac{1}{1-q} \sum_{k, s, R_1, R_2}^{\infty} k \, \mathcal{N}^{m,n,k}_{s, R_1, R_2} q^s \, \langle Tr_{R_1} U^n \rangle \langle Tr_{R_2} V^n \rangle
\eeq
The exponent disappeared since the partition function becomes trivial in the limit $m \to 0$.
The form of this expression explains why do we obtain knot polynomials in the expansion. However it does not explain why in the case at hand only the fundamental representations contribute and does not give us the coefficient in front of the polynomials. For that we have to know the BPS states degeneracies $\mathcal{N}^{\beta}_{s, R_1, R_2}$ explicitly.

After doing a geometric transition back to the $SU(2)$ picture we can express the coefficients of torus knot polynomials in terms of closed BPS states degeneracies. Indeed, the closed topological string free energy can be expressed as \cite{gopa}

\beq
F = \sum_{m,n,p,l,r,s} \sum_{j_L} \sum_{k=1}^{\infty} \frac{(-1)^{2 j_L} \mathcal{N}_{m,n, p, l, r, s}^{j_L}(q^{-2 k j_L}+ ... + q^{2 k j_L})}{k(q^{k/2}-q^{-k/2})^2} Q_{B}^{k m} Q_{F}^{k n} Q_{m1}^{k p} Q_{m2}^{k l} Q_{m3}^{k r} Q_{m4}^{k s}
\eeq
In the case of mass parameters $1$, $q$, $-A$, $Q_m$ the quantity we compute is just the derivative of it with respect to mass $m$ in the limit $m \to 0$  since the partition function goes to $1$ in this limit. If we consider the coefficient in front of $Q_B^m Q_F^n$ in this sum for coprime $(m,n)$ then there is only the contribution from the terms with $k=1$ and we obtain (up to an irrelevant factor)
\beq
H_{(m,n)} = \sum_{p,l,r,s} \sum_{j_L} \frac{(-1)^{2 j_L} \mathcal{N}_{m,n, p, l, r, s}^{j_L}(q^{-2 j_L}+ ... + q^{2 j_L})}{(q^{1/2}-q^{-1/2})^2} l (-A)^{p}  q^{r}
\eeq

So all the coefficients of the HOMFLY polynomial for torus knot $(m,n)$ can be expressed in terms of $\mathcal{N}_{m,n,p,l,r,s}^{j_L}$.

\subsection{View from IIB}

Let us consider the corresponding  counting problem in the IIB description. The particles are
represented by the string web involving the $ (p,q)$ strings. The rules
of interactions in the web and the attaching of the string web to the 5-brane web
are formulated in \cite{kol, sen} and it was argued that  there
are also the string strips corresponding to the strings located within 5 branes
and bound states of webs and strips when the strip escapes
from the hosting 5-brane. The knot polynomial in IIB counts the number
of the different string webs with the fixed boundary condition of the
string web at the 5-brane web. The counting of the spin content of the
particles corresponding to the web has been discussed in \cite{kol2} 
using the results from \cite{wittenphase}. It was shown that the spin 
content of the particles represented by web fits with the expected 
dimension of the moduli space \cite{wittenphase}.

Note that there is some subtle issue concerning
the role of the extended states in the physical space. 
The states with several quantum numbers can blow up and their
stability is supported by the angular momentum. In the IIB case the 
dyonic instanton
is represented by the D3 brane which can be considered as the blow-up 
of the string web. The $(n,m)$ quantum numbers are encoded in the 
particular solutions in the D3 brane worldvolume theory \cite{verlinde}. 
The total D3 brane charge should vanish hence it should have the closed
worldvolume and can be thought of as the $D3-\bar{D3}$ bound state
with the fixed electric and instanton charges $(p,q)$. 
It has the topology
of $T^2$ in $\mathbb{R}^4 \times S^1$ where radius of one circle equals to  $\beta$ 
while the second comes from the blow-up of the string-web and its 
radius equals to 
\beq
R^2 \propto nm
\eeq
The D3 brane do not shrink to the string due to the angular momentum
supported by the fields on its worldvolume.  In the physical space-time
it corresponds to the closed dyonic loop.

The HOMFLY polynomial $P_{n,m}(A,q)$ depends on two generating parameters 
$(A,q)$ and let us remind their   physical meaning . The q-parameter counts the angular 
momentum of the state in the self-dual $ \Omega$ background, while as was shown in \cite{gm}
the parameter $A$ counts the effects of the antifundamental mass. The string web with or 
without blow-up has evident similarities with 
the realization of HOMFLY as the weighted sum over the Dyck paths above the 
diagonal in the $(n,m)$ rectangle \cite{gor10}. The q-grading corresponds to the area under the path. 
If we assume web blow-up the  spin of the dyonic instanton is the product $nm$  and this product can be attributed 
to the boundary path without the corners. That is we conjecture  that the boundary path gets 
mapped to the dyonic instanton itself. Any other non-boundary paths involve corners and 
nontrivial counting with respect to the $A$-grading. We can conjecture that these paths
correspond to the generalization of the dyonic instantons in the theory with the
fundamental and antifundamental matter.

The question  if the blow up of the string web takes place and we could evaluate the knot
polynomial in terms of the D3 worldvolume theory deserves further study. This question 
is important for the issue of the interpretation of the HOMFLY purely in the $R^4\times S^1$ space-time
without appealing to the CY space.

\subsection{Analogy with the baryonic vertex}

Let us conclude this Section with  the conjecture that the hidden torus knot structure can be 
expected in the configuration
involving  the multiple Skyrmion charges in QCD. 
To formulate the conjecture  first remind the Skyrmion representation of the baryon found long
time ago in \cite{witten80}. The baryon was represented as the soliton state in
the Chiral Lagrangian and its fermion statistics is due to the 5d Chern-Simons
term.  The coefficient in front of the CS term equals to the number of colors
$N_c$.

In the  holographic approach 
the Chiral Lagrangian is the worldvolume theory on the flavor D8 branes. There
are two related ways to  describe  the baryon holographically. The baryonic vertex
can be represented by  the 5-brane wrapped around the compact cycle in the CY geometry
\cite{witten98}. The fundamental strings attached to the baryonic vertex are
extended along the radial coordinate and correspond to the electric degrees of
freedom at the boundary.

The baryon can be also represented as  the instanton 
in the $SU(N_f)_L\times SU(N_f)_R$  5d gauge theory on the flavor branes
\cite{son}.  The baryon-instanton is extended along the physical time. It is useful also  to
have  in mind the Atiyah-Manton representation of the Skyrmion from the
instanton holonomy  \cite{atiyah} when Skyrmion field is built from the components of the
instanton connection in $D=5$. The formal realization has been recognized in
terms of the instanton trapped inside the domain wall \cite{tong} in the
particular 5d gauge theory with a few flavors which has a finite number of
vacua.

The instanton interpretation  suggests the interesting  conjecture concerning the
possible place  of torus knots in the Skyrmion physics. The candidates for  the torus knot 
quantum numbers are evident 
and we can speculate that the conventional  charge B baryon  
at nonzero temperature is related  to the $T_{B, N_cB+k}$ torus
knot where the thermal circle plays the role of the KK circle in this case. The 
additional electric charge k can be related with the  F1 strings 
attached to the baryons-instantons. In the case of the flavor gauge group the corresponding 
electric strings correspond to the vector mesons. 
To pursue the analogy further we have to suggest some place for the entropic factor 
which follows from  degeneracy of the states with the fixed baryon and electric
quantum numbers $(B,3B+k)$.  

The analogy with the evaluation of the condensate goes as follows.  Consider the
chiral condensate which can be evaluated via the Casher--Banks relation in terms
of the Dirac operator spectrum. Consider the quark loop with inserted bilinear operator and 
additional  quark loop without the insertion. We could speculate that 
these two loops could be connected  by the baryon-meson web analogous to the
instanton-W-boson web in the SUSY case. The degeneracy in such web could play the
role similar to  the invariants.

One more remark is in order. In the SUSY case when we consider the torus knots with
coprime $(n,m)$ the instantons are sitting on the top of each other. However when we 
analyze the $(n,nk)$ torus links the centering at one point disappear and the
instantons form $n$ groups with $k$ instantons in each group \cite{etingof}. Hence if
we add just one unit of the electric charge to union of the $n$ links with linking number $lk_{i,j}=1$
the system gets topologically rearranged and become the single $(n,nk+1)$ torus knot.
If the analogy with the Skyrmion physics works it would mean that if we start with
the (B,3B) baryonic state and add the electric degree of freedom the state
gets rearranged and the multi-baryon state becomes the single $(B,3B+1)$ torus
knot.

Concluding this short comment on the analogy with Skyrmion physics   and chiral condensate 
in QCD  let us emphasize 
that the discussion  above was a bit speculative
however there are serious arguments to analyze this analogy further.

\section{Conclusion}

In this paper we discussed the role of the
knot invariants in the gauge theories and have  argued that the knot invariants count the entropy of the instanton--W-boson
web involved into the condensate formation in SQCD.
The picture is more transparent if the generic $T_{n,m}$ torus knots
are considered and it was shown that  the quantum numbers of the knot correspond to the 
instanton charge and
electric charge of 5D particles. The key point is that
the instanton--W-boson web  with  fixed two quantum numbers has  some entropic factor 
due to  the corresponding multiplicity which  is captured by the torus knot
invariants. We have seen such structure in the SU(2) SQCD or in simplified version of Abelian theory
supplemented by the particular Lagrangian branes.

During the consideration we have seen that there are two representations of the
HOMFLY invariants involving integer or  fractional 5d CS terms. This seems to
be parallel in many respects to the description of  FQHE via composite fermions
when the initial fermions identified as instantons get substituted by the
composite fermions with attached disorder analogous to the dyonic instantons in
our case. The relation with 4d and 2d  FQHE  seems to be deep and we shall
postpone this issue for  the separate publication.  We shall focus  at the
hydrodynamical aspects of the FQHE liquid which is substituted by the
hydrodynamical picture for the holomorphic instanton liquid \cite{hydro}.

In this paper we have focused at the 5d SUSY theory and made only  a short trip
to 4d theory when the knot invariants are encoded in the instanton
contributions in the corresponding theory. More detailed analysis of the
relation of the knot invariants with the different condensates in the 4d
theories is certainly required . Since the low-energy effective actions in 4d
theory in the Nekrasov-Shatashvili limit is governed by the quantum integrable
system it is very interesting to recognize the knot invariants in the  quantum integrable system. Another
interesting issue concerns the theories with less amount of SUSY when the
holomorphy is lost and instead of the instanton ensemble the
instanton--anti-instanton ensemble has to be investigated. In this case we could
expect a complicated linking phenomena responsible for the condensate
formation.

In \cite{bgn} we have observed the cascade of the different phase transitions 
in  the ensembles of instantons and the torus knots. Our consideration suggests the
possible stringy interpretation behind this phenomena. Indeed our evaluation of the condensate involves
the calculation of the different correlators of the Wilson loops or Wilson loops with the local
operators. When the number of W-bosons 
is large the instanton--W-boson web can be approximated by the surface and a kind of Gross-Ooguri 
phase transition  \cite{go} could take place which can be equivalently seen upon summation of the
ladder diagrams in the perturbation theory \cite{zarembo}.

Another interesting question concerns the application of the similar approach 
to the Schwinger process. At weak coupling in the worldline instanton formalism
we localize at the particular trajectories in the Euclidean space-time. Usually
one considers the single one loop $n$ the external field\cite{dunne}. However one could 
question about the role of two bounce configuration or additional local operator
apart from the bounce. The configuration of two bounces  involving different flavors is suppressed by the
additional exponential factor however twp loops could be related by the 
nontrivial instanton--W-boson web which produces the large entropic factor. The
large entropic factor from the web could also emerge when we consider the 
Euclidean loop and the separate local operator.

The interesting question concerns the fate of the  information stored
in this web upon the materialization of  Schwinger pairs after the Wick
rotation to the Minkowski space. It  could yield a
interesting entanglement factor. In the holographic picture the Schwinger
process requires the evaluation of the minimal surface \cite{gss}(see also
\cite{sz} for more recent discussion) probably with the additional operator insertion
if the effect of the condensate on the pair production is considered.
This corresponds to the
stable limit of the torus knots. When we consider two Euclidean circles connected by the web
 the saddle point solution gets modified. Let us also emphasize that in the unrefined 
case corresponding to the self-dual external field in some signature 
there is no Schwinger pair production however 
in the refined case these nonperturbative effects are unavoidable.

Concerning more formal problems we could mention first the generalization
to the colored HOMFLY polynomials. There are a few interesting questions
related to these issues. It is necessary to recognize the
physical interpretation of the  generating parameters
for the coloring. The natural candidate is the mass of the fundamental however
the immediate inspection shows that it can not be  literally true and more 
involved analysis is required. The knot polynomials with four gradings have been
considered in \cite{ggs}. From the instanton side in the colored case the centering
of the instantons at one point is destroyed and instantons are collected in the 
several groups located at the different points in $R^4$ \cite{etingof}. This suggests 
that the colored polynomials are related not to the condensates but to the 
topological correlators. 

This problem can be also considered for the generic
values of the masses of fundamentals and antifundamentals in SU(2) or higher 
rank  theory.
We expect that in the unrefined case the corresponding condensates are related to the combination 
of the products  of the HOMFLY  knot invariants. Indeed we have shown that each 
fermionic determinant dressed by $(n_i,m_i)$ numbers of  instantons and W-bosons
provides the HOMFLY invariant $P_{n_i,m_i}(q,a_i)$ where $a_i$ corresponds 
to the mass of the corresponding hyper. Hence the expected structure for the 
contribution in the $(n,m)$ sector is the product 
of the several HOMFLY invariants with the fixed value of the total instanton and 
electric charges.

The condensate  can be considered as the
derivative of the conformal block in the q-Liouville theory with respect to the 
parameter of the vertex operators. The interpretation
of the double expansion   of q-Liouville conformal block as the generating
function for the HOMFLY polynomials is quite promising and should clarify 
the way of regular evaluation of the knot invariants in terms of the 
2d conformal field theory.

Another immediate question concerns the recognition of the knot invariants
in the integrability framework using duality we have found. The 
relation with integrability  can be formulated, for instance,  in terms of the corresponding XXZ
spin chain and q-Liouville theory in the CY space or the Calogero model in the
physical space. The Whitham hierarchy should control the dependence of the
knot polynomials on the generating parameters. The several dualities known
in the integrability framework should be recognized and used in the knot
theory framework. As the simplest example remark that the $n\leftrightarrow m$
duality in the torus knot is the bispectral duality in the integrability framework.
The generalization to the $SU(N)$ case with the different matter is expected 
to provide more general knot invariants. We shall discuss these issues
elsewhere \cite{gmn}.

\section*{Acknowledgment}
We are grateful to  K.~Bulycheva, I.~Danilenko, E.~Gorsky, S.~Gukov, S.~Nechaev, 
N.~Nekrasov,  A.~Okounkov, Sh.~Shakirov,  A.~Vainshtein
and E.~Zenkevich  for useful
discussions.  The work of A.G. and A.M. was
supported in part by grants RFBR-15-02-02092    and Russian Science Foundation 14-050-00150 . A.G.
thanks the Simons Center for Geometry and Physics for the hospitality and support during the
program "Knot homologies, BPS states and Supersymmetric Gauge Theories". The work
of A.M. was also supported by the Dynasty fellowship program.

\appendix

\section{The refined topological vertex} {\label{app:vert}}

To establish some notations, let us very briefly review the topological vertex\cite{vafav, vafa07} calculations. In physical terms, 
topological vertexes compute 5D Nekrasov partition function for the gauge theory living on
a given web of $(p,q)$ 5-branes. In mathematical terms, they compute Gromov-Witten invariants for a given toric Calabi-Yau threefold. 

\begin{figure}[h]
\begin{center}
\setlength{\unitlength}{1.4cm}
\begin{picture}(3,3)
\linethickness{0.3mm}
\put(1,1){\line(1,0){1}}
\put(1,1){\line(0,1){1}}
\put(1,1){\line(-1,-1){0.7}}
\put(1.8,0.7){\makebox{$\nu$}}
\put(1.3,1.8){\makebox{$\mu$}}
\put(0.5,0.2){\makebox{$\lambda$}}
\end{picture}
\end{center}
\caption{The topological vertex.}
\label{fg:vert}
\end{figure}
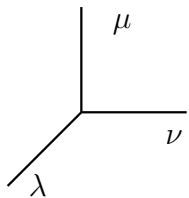

The building block is a 
trivalent vertex - Fig. \ref{fg:vert}.

In order to compute the partition function one has to divide the web into such vertexes, put a Young diagram on each internal line and empty Young diagram on
external lines and then sum over these diagrams. Each vertex contributes factor\footnote{Note that we use $1/q$ comparing to the original work \cite{vafa07}}
\beq
C_{\mu \nu \lambda}(q)= q^{-\frac{k(\nu)}{2}} s_\lambda(q^{\rho}) \sum_\tau s_{\mu^t/\tau}(q^{\rho+\la}) 
s_{\nu/\tau}(q^{\rho+\lambda^t}) 
\eeq
where $s_\la$ - Schur polynomials and $\rho_i=i-\frac{1}{2}$. We have used the following
functions on a Young diagram $\lambda$:
\begin{eqnarray}
||\la||^2=\sum_i \lambda^2_i \\
k(\lambda)=\sum_i \lambda^2_i-\lambda^{t \ 2}_i
\end{eqnarray}

Also, one has to take care of framing factors corresponding to internal lines\cite{vafa}. Without going into details, we just say that 
each internal line contributes
\beq
f_\nu(q)=(-1)^{|\nu|} q^{\frac{k(\nu)}{2}}
\eeq
to the power of line's framing.

Also, we can add Lagrangian branes to a toric diagram. From physical viewpoint they are D3 branes transversal to the original brane-web\footnote{Strictly speaking,
we are considering only Lagrangian branes projecting on the toric base as one dimensional submanifolds}. From mathematical viewpoint
they correspond to relative Gromov-Witten invariants relative to a Lagrangian submanifold in the CY. If we place a stack of branes on an external leg, then we have
to place a Young diagram $\mu$ on this leg and these Lagrangian branes contribute
\beq
s_\mu(-z_1,-z_2,\dots)
\eeq

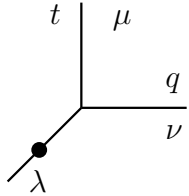
\begin{figure}[h]
\begin{center}
\setlength{\unitlength}{1.4cm}
\begin{picture}(3,3)
\linethickness{0.3mm}
\put(1,1){\line(1,0){1}}
\put(1,1){\line(0,1){1}}
\put(1,1){\line(-1,-1){0.7}}
\put(1.8,0.7){\makebox{$\nu$}}
\put(1.3,1.8){\makebox{$\mu$}}
\put(0.5,0.2){\makebox{$\lambda$}}
\put(1.8,1.2){\makebox{$q$}}
\put(0.7,1.8){\makebox{$t$}}
\put(0.6,0.6){\circle*{0.15}}
\end{picture}
\end{center}
\caption{The refined topological vertex. Dot indicates preferred direction.}
\label{fg:ref_vert}
\end{figure}

In \cite{vafa07}, Iqbal, Kozcaz and Vafa generalized this beautiful and powerful technique to general $\Om$-background:
\beq
C_{\mu \nu \lambda}(t,q)= (q t)^{-\frac{||\nu||^2+||\lambda||^2}{2}} t^{\frac{k(\nu)}{2}} P_\lambda(t^{-\rho},1/q,t) 
\sum_\tau (q t)^{\frac{|\nu|-|\tau|-|\mu|}{2}}  s_{\mu^t/\tau}(t^{-\rho} q^\la ) 
s_{\nu/\tau}(q^{\rho} t^{-\lambda^t}) 
\eeq
where $P_\la$ is MacDonald polynomial and
\begin{eqnarray}
q=\exp(-\beta \ep_1) \\ \nonumber
t=\exp(-\beta \ep_2) 
\end{eqnarray}
And the framing factor reads as:
\beq
f_\nu(t,q)=(-1)^{|\nu|} (tq)^{\frac{|\nu|-||\nu||^2}{2}} q^{+\frac{k(\nu)}{2}}
\eeq
Now the vertex has no cyclic symmetry, and one has to choose preferred direction(see Fig. \ref{fg:ref_vert}). Usually, the answer
does not depend on the particular choice for closed amplitudes, but it is not always so for open amplitudes\cite{vafa07,ak:change}.

For all known examples it reproduces Nekrasov instanton formulas. However, there are still a plethora of unaswered 
questions. 
For example, even before \cite{vafa07}, in \cite{ak:vert} Awata and Kanno proposed another version of refined vertex. 
Again, for all known closed amplitudes the Iqbal-Kozcaz-Vafa(IKV) and Awata-Kanno(AK) vertexes give the same answer. Nonetheless,
we will use the IKV vertex with caution. 

\section{Chiral ring and Lagrangian branes} \label{app:br}

Following \cite{nekrasov12} we will introduce generating function $Y(z)$
\beq
Y(z)=\exp \l \sum_{n=1} \cfrac{z^{-n}}{n} \mathcal{O}_n \r
\eeq
for expectation values of chiral ring operators:
\beq
\mathcal{O}_n=\ll \exp(-n \beta \Phi) \rr
\eeq

One can show that in terms of instanton partitions $Y(z)$ reads as\cite{nekrasov14}:
\beq
Y(z)=\cfrac{1}{(1-z)} \cfrac{\prod_{\square \in \pr_+ \la}(z- q^{a'} t^{l'})}{\prod_{\square \in \pr_- \la}(z- q t  q^{a'} t^{l'})}
\eeq
Let us show that this function is actually a wave-function for brane-antibrane system.
We have seen that a single Lagrangian brane with zero framing on an external leg contributes factor

\beq
M(z) = \prod_{j=1}^{l(\la)} \cfrac{1-z t^{j-1} q^{\la_j}}{1-z t^{j-1}} 
\eeq

to the $U(1)$ instanton partition function.

Also, depending on the leg and framing we will arrive at either $M(z)$ - for brane,  or $1/M(z)$ - for antibrane.

\begin{figure}[h]

\begin{center}

\setlength{\unitlength}{1.0cm}

\begin{picture}(8,4)

\linethickness{0.3mm}

\put(2,0){\line(1,1){1}}

\put(3,1){\line(1,0){1}}

\put(4,1){\line(1,-1){1}}

\put(3,1){\line(0,1){1}}

\put(4,1){\line(0,1){1}}







\put(2.5,0){\vector(-1,1){0.7}}

\put(4.5,0){\vector(1,1){0.7}}

\put(2.5,0.3){\makebox{$z_1 \sqrt{q}$}}

\put(5.3,0.3){\makebox{$z_2 \sqrt{q}$}}

\end{picture}

\end{center}

\caption{$U(1)$ theory with two Lagrangian branes on external legs. We should choose zero framing for the left brane and $+1$
for the right one}

\label{fg:su3}

\end{figure}
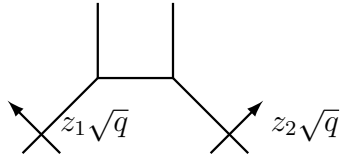

Actually, functions $Y(z)$ and $M(z)$ are not independent: they obey a very simple relation:

\beq
\label{eq:y_m}
Y(z) = \cfrac{M(1/z)}{M(1/zt)}
\eeq	

The proof consists of a simple comparing which boxes in Young diagrams actually contribute the left hand side and
right hand side.

To obtain a ratio of two $M$ functions, we can consider 
two Lagrangian branes(Fig. \ref{fg:su3}). 

\beq
Z=\sum_{\la, \mu, \nu} (-Q)^{|\la|} C_{\emptyset \nu \la}(q,t) C_{\mu \emptyset \la^t}(t,q) f_\la(t,q) f_\nu(t,q) s_\nu(-z_2 \sqrt{q}) s_\mu(-z_1 \sqrt{q})
\eeq

Lagrangian branes contribute:
\beq
\cfrac{M(z_1)}{M(z_2)}
\eeq
For $z_2=z_1 t=z t $ we obtain exactly instanton part of $Y(1/z)$.

Since $Y(z)$ is the generation function for chiral ring operators,
we conclude that these operators could be obtained by a brane-antibrane lump of size $\ep_2$.

Actually, we can invert (\ref{eq:y_m}):
\beq
M(z)=\prod_{i=0}^\infty Y(z^{-1}t^{-i})
\eeq
Or in terms of chiral ring VEVs:
\beq
M(z)=\exp \l \sum_{n=1}^\infty \cfrac{z^n t^n \mathcal{O}_n}{n(t^n-1)} \r
\eeq
Actually it is easy to generalize the above formulas to the $SU(N)$ case but we postpone this to the future work.

\section{Jones-Rosso formula} \label{app:a}
The Jones-Rosso formula for the HOMFLY-PT polynomial of $(n,m)$ torus knot colored in the representation $R$ reads as
follows:
\beq
\label{jr}
H^{(n,m)}_R(A,q)=\mathlarger{ \sum_{\la \in R^{\otimes n}} q^{\frac{m}{n} \sum_{\square \in \la}(a - l)} 
c_R^\la \chi_\la(p^*)}
\eeq
where:
\begin{itemize}
\item $R$ - Young diagram defining the representation
\item $c_R^\la$ - Adams coefficients, defined by the action of the Adams operation on Schur polynomials $\chi_\mu(p)$:
\beq
\chi_\mu(p^{(n)})=\sum_{\eta \in \mu^{\otimes n}} c_\mu^\eta \chi_\eta(p)
\eeq
In our notation we write arguments of Schur polynomials as \textit{power series} polynomials $p_k=x_1^k+x_2^k+...$ and
\beq
p^{(n)}_k=p_{nk}=x_1^{nk}+x_2^{nk}+...
\eeq
\item Finally, $p^*$ define the special choice of power series polynomials:
\beq
\label{times}
p_k^*=\cfrac{(-A)^k-(-A)^{-k}}{q^k-q^{-k}}
\eeq
\end{itemize}

Lets rewrite the Jones-Rosso formula (\ref{jr}) in a more explicit form. It is very-well known\cite{macdonald} 
that for the special choice of $p^*$ (\ref{times}), Schur polynomials read as:
\beq
\chi_\la(p^*)=\mathlarger{ q^{\sum a} \cfrac{\prod (1+A q^{l'-a'})}{\prod(1-q^{a+l+1})} }
\eeq
If we confine ourselves to the fundamental representation $R=\square$ then it is possible to obtain an 
explicit expression for the Adams coefficient\cite{shamil}:
\beq
c_\square^\la = \mathlarger{ q^{\sum a} (1-q^n) \cfrac{\prod^{0,0}(1-q^{l'-a'})}{\prod(1-q^{a+l+1})} }  
\eeq

Combining all the factors we obtain the following expression for the HOMFLY-PT polynomial in fundamental representation
for $(n,m)$ torus knot(we omitted the trivial factor $(1+A)$):
\beq
\label{homfl1}
\homf=(1-q^n) \sum_{|\la|=n} \mathlarger{ q^{2 \sum a}  q^{\frac{m}{n}\sum(a-l)}
\cfrac{\prod^{0,0}(1-q^{l'-a'}) \prod^{0,0}(1+A q^{l'-a'})}{\prod(1-q^{a+l+1})^2} }
\eeq
Or equivalently:
\beq
H^{(n,m)}_\square(A,q)=(-1)^{n-1}\cfrac{1-q^n}{q^n} \sum_{|\la|=n} \mathlarger{  q^{(\frac{m}{n}+1)\sum(a-l)}
\cfrac{\prod^{0,0}(1-q^{a'-l'}) \prod^{0,0}(1+A q^{l'-a'})}{\prod(q^{-l-1}-q^a)(q^{-l}-q^{a+1})} }
\eeq

\section{Lagrangian brane and JR formula} \label{app:b}
Now let us show that the formula (\ref{new}) in the unrefined case does indeed reproduce the JR expression.
First of all, note that because of the factor $\prod^{0,0}(1-q^{l'-a'})$ only hook-shaped Young diagrams contribute.

\begin{figure}[h]
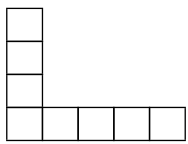

\begin{equation}
\begin{Young}
\cr
\cr
\cr 
&&&&\cr
\end{Young}
\nonumber
\end{equation}
\caption{Hook-shaped Young diagram}
\end{figure}
Suppose that the hook-shaped diagram $\la$ has the horizontal "arm" of length $w$. Then the vertical "leg" has length $n-w+1$,
since the total number of boxes is $n$. Chern-Simons term in the JR formula gives
\beq
q^{\frac{m}{n} \l \sum a- \sum l \r}=q^{m w - \frac{m(n+1)}{2}}
\eeq
Whereas the contribution from Lagrangian brane:
\beq
M(z)=\cfrac{1-q^w z }{1-q^{w-n} z}
\eeq
Therefore
\beq
Coef_{z^m} M(z) = q^{m(w-n)}(1-q^n)
\eeq
We see that apart from the normalization factor $q^{m(1-n)/2}$ these two expressions coincide.

\printbibliography

\end{document}